\documentclass[journal]{IEEEtran}
\usepackage{eurosym}
\usepackage{microtype}
\usepackage{amssymb}
\usepackage{yfonts}
\usepackage{cite}
\usepackage{url}
\usepackage{multirow}
\usepackage{xcolor}
\usepackage{blindtext}
\usepackage{eucal}
\usepackage{enumitem}
\usepackage{amsmath}
\usepackage{algorithmic}
\usepackage{tikz}
\newcommand*\circled[1]{\tikz[baseline=(char.base)]{
            \node[shape=circle,draw,inner sep=1.5pt] (char) {#1};}}

\usepackage{nomencl}
\usepackage{etoolbox}
\makenomenclature

\renewcommand\nomgroup[1]{%
\color{black}
  \item[\bfseries
  \ifstrequal{#1}{P}{Variables and functions}{%
  \ifstrequal{#1}{N}{Constants}{%
  \ifstrequal{#1}{O}{Sets and Indices}{}}}%
]\vspace{7pt}}

\usepackage{amsfonts}

\usepackage{array}

\usepackage[caption=false]{subfig}

\makeatletter
\newcommand{\thickhline}{%
    \noalign {\ifnum 0=`}\fi \hrule height 0.75pt
    \futurelet \reserved@a \@xhline
}
\newcolumntype{"}{@{\hskip\tabcolsep\vrule width 0.75pt\hskip\tabcolsep}}
\makeatother

\begin{document}

\title{Strategic Scheduling of Discrete Control Devices in Active Distribution Systems}

\author{Alireza~Nouri,~\IEEEmembership{Member,~IEEE,}
Alireza~Soroudi,~\IEEEmembership{Senior~Member,~IEEE,}
and~Andrew~Keane,~\IEEEmembership{Senior~Member,~IEEE}
\thanks{A. Nouri (alireza.nouri@ucd.ie), A. Soroudi (alireza.soroudi@ucd.ie) and A. Keane (andrew.keane@ucd.ie) are with the School of Electrical and Electronic Engineering, University College Dublin, Dublin 04, Ireland.

This work has emanated from research conducted with the financial support of Science Foundation Ireland under the SFI Strategic Partnership Programme Grant Number SFI/15/SPP/E3125. The opinions, findings and conclusions or recommendations expressed in this material are those of the authors and do not necessarily reflect the views of the Science Foundation Ireland.}
}

\markboth{}%
{Shell \MakeLowercase{\textit{et al.}}: Bare Demo of IEEEtran.cls for Journals}
\maketitle

\begin{abstract}
\textcolor[rgb]{0,0,0}{The frequent actuation of discrete control devices (\textsc{dcd}s), e.g., on-load tap changers, drastically reduces their lifetime. This, in turn, imposes a huge replacement cost.
Simultaneous scheduling of these \textsc{dcd}s and continuous control devices, e.g., distributed energy resources, is imperative for reducing the operating cost. This also increases the lifetime of \textsc{dcd}s and helps to avoid the sub-optimal/infeasible solutions.
Considering the high cost of discrete control actions (\textsc{dca}s), they may never be justified against the other options in a short scheduling horizon (\textsc{sh}).
With a longer \textsc{sh}, their benefits over a long period justify \textsc{dca}s.
However, a shorter \textsc{sh} helps to hedge against the risk impelled by uncertainties.
Here, the system future is modeled as a set of multi-period scenarios. The operator exploits a long \textsc{sh}, but solely applies the decisions made for the first period and waits for updated data to make the next decisions.}
This enables cost reduction by strategically applying \textsc{dca}s prior to the time that they are inevitable, while avoiding them when unneeded.
\textcolor[rgb]{0,0,0}{The proposed branch-and-cut-based solution methodology} accurately deals with \textsc{dca}s while applying some expediting heuristics.
During the branching process, a globally convergent trust region algorithm solves the integer relaxed problems.
\end{abstract}
\IEEEpeerreviewmaketitle

\nomenclature[P, 01]{$CC_t$}{Cost of discrete control actions}
\nomenclature[P, 03]{$I_l$}{Current of line $l$}
\nomenclature[P, 04]{$I_p/I_s$}{Current of primary/secondary winding of transformer}
\nomenclature[P, 05]{$I^*$}{Conjugate of vector $I$}
\nomenclature[P, 06]{${P_d}_b$}{Voltage dependent active power demand at bus $b$}
\nomenclature[P, 07]{${P_g/Q_g}$}{Active/reactive power generated by continuous devices}
\nomenclature[P, 08]{$P_p/Q_p$}{Active/reactive power injected by upstream network}
\nomenclature[P, 09]{${Q_d}_b$}{Voltage dependent reactive power demand at bus $b$}
\nomenclature[P, 12]{$r_{k_{TC},t}$}{Turn ratio of transformer $t$ (pu)}
\nomenclature[P, 13]{$st_{k_{CB}}$}{Step of \textsc{cb} $k_{CB}$ which is inserted into circuit}
\nomenclature[P, 14]{$tap_{k_{TC}}$}{Tap position of \textsc{oltc} $k_{TC}$}
\nomenclature[P, 15]{$u_{k_{CB},t}$}{Binary variable indicating if the step of \textsc{cb} $k_{CB}$ is changed. 1 if changed.}
\nomenclature[P, 16]{$u_{k_{TC},t}$}{Binary variable indicating if the tap position of \textsc{oltc} $k_{TC}$ is changed. 1 if changed.}
\nomenclature[P, 18]{$V_b$}{Voltage phasor at bus $b$}
\nomenclature[P, 19]{$V_x/V_y$}{Real/Imaginary part of phasor $V$}
\nomenclature[P, 21]{$\left|V_b\right|$}{Magnitude of voltage at bus $b$}
\nomenclature[P, 22]{$V_p/V_s$}{Voltage at primary/secondary side of transformer}
\nomenclature[P, 23]{${VD_b}_{t,s}$}{Voltage deviation at bus $b$ in period $t$ and scenario $s$}
\nomenclature[P, 24]{$Z_p$/$Y_p$}{Parallel impedance/admittance at the primary side of equivalent $\pi$ model as a function of $tap$}
\nomenclature[P, 25]{$Z_s$/$Y_s$}{Parallel impedance/admittance at the secondary side of equivalent $\pi$ model as a function of $tap$}
\nomenclature[P, 26]{$Z_{sr}$}{Series impedance in equivalent $\pi$ model}
\nomenclature[P, 26]{$Y_{sr}$}{Series admittance in equivalent $\pi$ model}
\nomenclature[P, 27]{$\delta_V$}{Angle of phasor V}
\nomenclature[P, 28]{$\Delta \upsilon_{t,s}$}{Perturbed value of control variable $\upsilon$, i.e., $\upsilon_{t,s}$-$\hat{\upsilon}_{t,s}$}
\nomenclature[P, 29]{$\Delta OC_t$}{Change in the operation cost of period $t$ ($OC_t$-$\hat{OC}_t$)}
\nomenclature[P, 30]{$\gamma_m$}{Linear counterpart of $\nu_m {V_s}_y$}
\nomenclature[P, 31]{$\varphi_{I,V}$}{Phase difference between $I$ and $V$}
\nomenclature[P, 32]{$\nu_m$}{Dummy binary variables to extract \textsc{oltc} linear model}
\nomenclature[P, 33]{$\underline{{\vartheta_b}_{t,s}}$}{Under-voltage magnitude at bus $b$ in period $t$ and scenario $s$}
\nomenclature[P, 34]{$\overline{{\vartheta_b}_{t,s}}$}{Over-voltage magnitude at bus $b$ in period $t$ and scenario $s$}
\nomenclature[P, 35]{$\sigma$}{Binary variable in order to extract the originally linear model of \textsc{oltc} transformer. If ${V_s}_y \geq 0$, $\sigma$=1.}
\nomenclature[P, 36]{$\upsilon_{t,s}$}{Optimization variable $\upsilon$, in period $t$ and scenario $s$}
\nomenclature[P, 37]{$\upsilon_{\tau+1}$}{Optimization variable $\upsilon$, in period $\tau+1$ which is independent of scenario $s$}
\nomenclature[P, 39]{$\Upsilon$}{Vector of optimization variables, e.g., $Pg$}
\nomenclature[P, 40]{$\underline{\epsilon_b}/\overline{\epsilon_b}$}{Auxiliary variable to change the minimum/maximum voltage constraints at bus $b$ to an equality constraint}

\nomenclature[N, 01]{${a_p}_b$}{Constant impedance share in ZIP load model at bus $b$}
\nomenclature[N, 02]{${a'_p}_b$}{Constant impedance share in ZP load model at bus $b$}
\nomenclature[N, 03]{${b_p}_b$}{Constant current share in ZIP load model at bus $b$}
\nomenclature[N, 04]{${c_p}_b$/${c'_p}_b$}{Constant power share in ZIP/ZP load model at bus $b$}
\nomenclature[N, 06]{$\textsl{D}_{k_{TC}}$}{Tap-changing operation cost of \textsc{oltc} transformer $k_{TC}$}
\nomenclature[N, 07]{$\textsl{D}_{k_{CB}}$}{Step-changing operation cost fof \textsc{cb} $k_{CB}$}
\nomenclature[N, 08]{$e_{t,\xi}$}{Standard deviation of parameter $\xi$ in period $t$}
\nomenclature[N, 09]{$\overline{G}/\underline{G}$}{Maximum/minimum value of variable $G$}
\nomenclature[N, 10]{$P_{d_0}$}{Active power demand at voltage $V_0$}
\nomenclature[N, 12]{$Q_{d_0}$}{Reactive power demand at voltage $V_0$}
\nomenclature[N, 13]{$\overline{Q_g^{wind}}$}{Minimum reactive power injection of wind generators}
\nomenclature[N, 14]{$R_c^n$}{Nominal transformer core resistance (pu.)}
\nomenclature[N, 15]{$R_l$/$X_l$}{Resistance/reactance of line $l$}
\nomenclature[N, 16]{$T$}{Scheduling time horizon in hour}
\nomenclature[N, 17]{${V_0}_b$}{Nominal voltage at bus $b$}
\nomenclature[N, 18]{$V_{th}$}{Upstream system Thevenin voltage}
\nomenclature[N, 19]{$X_M^n$}{Nominal transformer magnetizing reactance (pu.)}
\nomenclature[N, 20]{$Y_\text{Bus}$}{Network admittance matrix}
\nomenclature[N, 21]{$y^{st}$}{\textsc{cb} admittance step change}
\nomenclature[N, 22]{$Z^n_{sr}$}{Nominal series impedance of transformer $t$ (pu.)}
\nomenclature[N, 23]{$Z^n_{sr,p}$}{Nominal transformer primary series impedance (pu.)}
\nomenclature[N, 24]{$Z^n_{sr,s}$}{Nominal transformer secondary series impedance (pu.)}
\nomenclature[N, 25]{$Z_{th}$}{Upstream system Thevenin impedance}
\nomenclature[N, 26]{$\mu_{t,\xi}$}{Average values of uncertain parameter $\xi$ in period $t$}
\nomenclature[N, 27]{$\pi_s$}{Probability of scenario $s$}
\nomenclature[N, 28]{$\overline{\alpha^{pv}}$}{Maximum power angle of photovoltaic units}
\nomenclature[N, 29]{$\eta_b$}{A large positive number signifying the importance level of voltage control at bus $b$}
\nomenclature[N, 30]{$\theta^{der}$}{Maximum power angle allowed for \textsc{der} $der$}
\nomenclature[N, 31]{$\rho_A/\rho_R$}{Active/reactive power consumption price}
\nomenclature[N, 32]{$\rho_e$}{Energy price of \textsc{der} $e$}
\nomenclature[N, 33]{$\sigma_t$}{Time period $t$ in hour}
\nomenclature[N, 34]{$\tau$}{First time period of the previous scheduling window}
\nomenclature[N, 35]{$\hat{\Upsilon}$}{Vector of optimization variables found in the previous \textsc{tra} sub-problem or vector of initial solution}
\nomenclature[N, 36]{${\Delta}U$}{Voltage step change per tap in pu}

\nomenclature[O, 02]{$N_b$}{Number of of buses indexed by $b$}
\nomenclature[O, 03]{$N_m$}{Number of tap positions indexed by $m$}
\nomenclature[O, 04]{$N_m^-$}{Number of negative tap positions indexed by $m$}
\nomenclature[O, 05]{$N_s$}{Number of scenarios indexed by $s$}
\nomenclature[O, 06]{$N_\upsilon$}{Number of control variables}
\nomenclature[O, 07]{$\overline{N_{TC_T}}$}{Maximum number of tap changing actions for \textsc{sh} $T$}

\vspace{-1mm}
\printnomenclature
\vspace{-1mm}
\section*{\textcolor[rgb]{0,0,0}{List of Abbreviations}}
\addcontentsline{toc}{section}{List of Abbreviations}
\begin{IEEEdescription}[\IEEEusemathlabelsep\IEEEsetlabelwidth{AAAa}]
\color{black}
\item[\textsc{bc}]    Branch and cut
\item[\textsc{cb}]    Capacitor bank
\item[\textsc{dca}]   Discrete controllable action
\item[\textsc{dcc}]   Discrete control cost
\item[\textsc{dcd}]   Discrete controllable device
\item[\textsc{der}]   Distributed energy resource
\item[\textsc{dsp}]   Distribution scheduling problem
\item[\textsc{lp}]    Linear programming
\item[\textsc{milp}] Mixed integer linear programming
\item[\textsc{minlp}] Mixed integer nonlinear programming
\item[\textsc{nlp}]   Nonlinear programming
\item[\textsc{oc}]    Operation cost
\item[\textsc{oltc}]  On load tap changer
\item[\textsc{pdf}]   Probability density function
\item[\textsc{res}]   Renewable energy sources
\item[\textsc{rhc}]   Receding horizon control
\item[\textsc{sh}]    Scheduling horizon
\item[\textsc{socp}]  Second-order cone programming
\item[\textsc{svr}]   Static voltage regulator
\item[\textsc{sa}]    Solution approaches
\item[\textsc{tra}]   Trust region algorithm
\end{IEEEdescription}

\vspace{-3pt}
\section{Introduction}\label{sec_1}
\IEEEPARstart{W}{ith} the increasing penetration of highly variable renewable production in active distribution systems, the resource scheduling and voltage control problem is increasingly becoming dynamic in nature. \textcolor[rgb]{0,0,0}{The variability of load levels and upstream system characteristics give rise to this issue.}
\textcolor[rgb]{0,0,0}{Discrete control devices (\textsc{dcd}s), e.g., on load tap changers (\textsc{oltc}s) and capacitor banks (\textsc{cb}s), are able to control the voltage levels in the steady state conditions.
However, maintaining the voltage levels within the predefined limits with only conventional \textsc{dcd}s entails frequent actuation of these devices. This reduces their lifetime. }
Due to their high cost, discrete control actions (\textsc{dca}s), cannot be applied frequently. The use of \textsc{dcd}s should be optimally balanced with continuously controllable devices, e.g., distributed energy resources (\textsc{der}s).
\textcolor[rgb]{0,0,0}{In a short scheduling horizon (\textsc{sh}), \textsc{dca}s may never be justified against the cheaper changes in active and reactive power supports of the continuous control devices. Therefore, they would never be applied unless they are inevitable.
With a longer \textsc{sh}, their higher impact on operating cost reduction may justify these actions.}
In previous studies, \textsc{dca}s were usually scheduled prior to the real-time operation, e.g., in a day-ahead period \cite{Xu2017}.
Conversely, to reduce the effects of uncertainties and to keep the solution optimality,
the \textsc{sh} should be as short as possible. Therefore, each short and long \textsc{sh} has its own cons and pros.

\textcolor[rgb]{0,0,0}{Here, the term ``Discrete Control Devices'' is referred to a family of equipment and devices with a discrete controllable parameter.
The main focus is on the co-optimization of the schedule of discrete and continuous control devices and steady state voltage control.
To this end,} the system future is modeled as a set of scenarios.
Each scenario includes load levels, load characteristics,
production levels of renewable energy sources (\textsc{res}s), upstream system characteristics and utility prices, all in successive periods of the \textsc{sh}.
In the first period, the same decisions are taken for all scenarios.
Only the decisions taken for this period are implemented. The framework will be applied again to find the optimal schedule of the next periods.
In this way, the decision maker benefits a long \textsc{sh}, makes the decisions based on the available
and forecast data and waits for more accurate data to make the next decisions.
Some \textsc{dca}s are inevitable in the upcoming periods.
Using the proposed method, the operator may be able to reduce the cost by applying these \textsc{dca}s in advance.
\textcolor[rgb]{0,0,0}{Since only the decisions of the first period are applied, the effects of uncertainties are also restricted.}

The proposed solution methodology based on branch and cut (\textsc{bc}) technique accurately deals with the discrete control variables.
This helps to avoid infeasibility/sub-optimality.
\textcolor[rgb]{0,0,0}{During the branching process of branch and cut technique, an integer relaxed problem with additional bounds on integer variables should be solved at each node.}
A globally convergent trust region algorithm (\textsc{tra}) is applied to solve these integer relaxed problems.
\textcolor[rgb]{0,0,0}{The \textsc{tra} sub-problems are solved using interior point method \cite{BAHRAMI2020}.
The details of the proposed solution methodology are provided in subsection \ref{Sol}.}
The proposed method incorporates the accurate models and does not rely on simplifying assumptions
such as those made in convex relaxation approaches, e.g., balanced operation (see \ref{LitRev}).
Upstream system, \textsc{oltc}, \textsc{cb}s and voltage dependent loads are accurately modeled.
These models are kept up-to-date while solving the distribution scheduling problem (\textsc{dsp}).
\vspace{-2.7mm}
\subsection{Literature Review}\label{LitRev}
A multi-timescale coordinated control was proposed \cite{Xu2017} and \cite{Zheng2017}.
\textcolor[rgb]{0,0,0}{The system uncertainties were considered in the day-ahead scheduling of \textsc{dcs}s. However, there is still no guarantee that the fast (continuous) control devices can maintain the voltages within the limits in the intra-day scheduling problems.}
Moreover, during the next day, the operator is able to further reduce the system cost by readjusting the schedule of the \textsc{dcd}s
based on the uncertainties already revealed.
The main focus of \cite{OConnell2017} was on presenting a comprehensive framework for uncertainty handling in an unbalanced three-phase system based on information gap decision theory.
A day-ahead scheduling was solved for \textsc{dca}s. 
In this paper, the schedule of \textsc{dca}s is changed according to as much data as available in the real-time operation and the latest forecast on the uncertain parameters based on receding horizon control (\textsc{rhc}).
Tap-changing cost was modeled in \cite{Ren2017}. The same model is used in this paper.

\textsc{rhc} technique was implemented in distribution systems to achieve different goals. This technique deals with the constrained dynamic optimization problems by
finding a set of consecutive control actions derived by optimizing the objective function over a horizon window.
\textsc{rhc} was applied in \cite{RHCDG} to find the optimal set-points of \textsc{der}s, in \cite{Rolling}
to deal with the dynamics of energy storage systems and in \cite{RHCDefer} to schedule the deferrable loads.
\textsc{dcd}s were not included in these studies
The ability of \textsc{rhc} technique
to solve the issues related to the scheduling of \textsc{dcd}s has not yet been analyzed.
With this technique employed to address the aforementioned issues of \textsc{dcd}s in AC \textsc{dsp}, computational burden at each time step can be deemed as a downside.
The proposed method exploits the \textsc{rhc} technique to handle the \textsc{dcd}s and continuous control devices simultaneously.
The computational burden is dealt with the solution methodology proposed in subsection \ref{Sol}.

\textcolor[rgb]{0,0,0}{It was shown in \cite{KHALILI201992} how the energy storage systems and demand response programs mitigate the effects of uncertainties.
The main focus was on the planning of the battery storage systems and the \textsc{dcd}s were neglected.
The operation sub-problems were also solved to achieve the feasible and optimal solutions.
The storage systems can also be modeled in the formulation proposed in the present paper based on \cite{Rolling} and considering the practical requirements presented in \cite{KHALILI201992}.
Most demand response programs and reliability measures cannot be modeled in a short \textsc{sh}. The proposed \textsc{rhc}-based scheduling can also be applied to handle these programs. The practical consideration of demand response programs and reliability concerns were discussed in \cite{KHALILI2019429} and \cite{Khalili2019}, respectively.}

The \textsc{dsp} is a \textsc{minlp} problem
that takes a long time to converge to a reliable solution. The uncertainties escalates the issue.
Some previous works applied simplified formulations, e.g., linearized formulations \cite{Linear},
and also simplified models for \textsc{dcd}s, e.g., round-off approaches \cite{RoundOff}.
There is no guarantee that the solution is optimal or even feasible with these methods.

The state-of-the-art second order cone programming
(\textsc{socp}) and the semidefinite programming based on branch flow model
\cite{Li2018}
have gained popularity in recent years.
These methods are able to effectively solve the \textsc{dsp} in balanced active distribution systems
enabled with continuous control devices under quite acceptable assumptions \cite{McCormick}.
\textsc{socp} was applied in \cite{Stoch} to solve the stochastic day-ahead \textsc{dsp}.
\textsc{dcd}s, e.g., \textsc{oltc}s and \textsc{cb}s, was also taken into account. In order to model \textsc{oltc} transformers, it was assumed that
the primary voltage is fixed to avoid the non-linearity introduced by the discrete-linear product terms of turn ratio
and primary voltage. The effects of bus voltage on the reactive power injected by \textsc{cb}s was also neglected for the same reason.
Additional binary and continuous variables were introduced in \cite{LinearOLTC} to present a linearized model for \textsc{oltc}s.
\textsc{cb}s can be modeled in the same way. For a long \textsc{sh} (which is inevitable with \textsc{dcd}s),
the number of binary variables will drastically increase leading to a long solution time.
In \cite{McCormick}, a McCormick relaxation was employed within a sequential bound-tightening algorithm to tackle the problem.
For an unbalanced active distribution systems the convex relaxation techniques based on branch flow model cannot be applied to effectively
solve the \textsc{dsp} under acceptable assumptions.
semidefinite programming was used in \cite{UnbalancedAlaki} to solve the \textsc{dsp} in an unbalanced system without considering the mutual inductance between the phases.
Neglecting the voltage unbalance at system buses, semidefinite programming was applied in \cite{RoundOff} and \cite{Unbalanced} to solve the \textsc{dsp}.
These assumptions are not reasonable for practical systems, where voltage unbalance really matters and sometimes voltage unbalance minimization is at least considered as one of the objectives. 
\vspace{-3mm}
\subsection{\textcolor[rgb]{0,0,0}{Novelty and Contributions}}\label{Contrib}
\begin{enumerate}
\item to exploit the advantages of both short and long \textsc{sh}s using \textsc{rhc} technique
which enables \textit{short-term co-optimization} of the slow and fast control devices considering the system \textsl{uncertainties}.
\item to keep the upstream and load models up-to-date while solving the \textsc{dsp}.
\item to tailor the stochastic scheduling formulation so that the integer relaxed problems can be solved using \textsc{tra}.
\item to compare the efficiency of the perturbed models of \textsc{dcd}s and their originally linear models developed using auxiliary binary variables.
\end{enumerate}
\vspace{-3mm}
\subsection{\textcolor[rgb]{0,0,0}{Paper organization}}\label{Contrib2}
\textcolor[rgb]{0,0,0}{An overview of the proposed scheduling algorithm is presented in Section \ref{Prop}.
Section \ref{Deter_Alg} presents the deterministic \textsc{dsp} formulation.
Section \ref{Stoch_Alg} shows how to extend this formulation to include the effects of uncertainties.
Several studies are designed in Section \ref{Case_Studies} to test the performance of the proposed method. 
The conclusions are drawn and directions for future studies are outlined in Section \ref{conclusion}.}
\vspace{-1pt}
\section{Proposed Methodology}\label{Prop}
\subsection{Uncertainty Modeling}\label{Uncertain}
\textcolor[rgb]{0,0,0}{In \textsc{rhc} technique, the set of scenarios of each scheduling window
should be updated using the latest update on the input data.} For each scheduling window, scenario generation is an iterative process along the successive periods of this window.
Starting from node $b_t$ at stage $t$  ($\hspace{1mm} \tau \leq t \leq \tau+T-1$), first a joint probability density function (\textsc{pdf}) is developed for the uncertain parameters at stage $t$+1.
\textcolor[rgb]{0,0,0}{To form such a joint \textsc{pdf}, the parameters of this \textsc{pdf}, i.e., statistical moments such as mean values and variances, are extracted from the historical data available up until time period $t$.}
Then a single stage set of scenarios is developed using the method presented in \cite{Daobao2012} based on moment-matching technique.
K-means clustering technique \cite{Daobao2012} is applied to reduce the number of single stage scenarios to keep the scenario generation tractable.

\textcolor[rgb]{0,0,0}{The details of developing the single stage scenarios using \textsc{lp} moment-matching and K-means clustering techniques were best explained in \cite{Daobao2012}.
Here the process is outlined briefly.
The objective in \textsc{lp} moment-matching technique is to find a set of scenarios for which the statistical moments are as much close as possible to the statistical moments of the joint \textsc{pdf} of the uncertain parameters.
A large number of single stage scenarios are first generated for all uncertain parameters.
Then, the probability of each scenario is determined using \textsc{lp} technique so that the objective of \textsc{lp} moment-matching technique is satisfied as much as possible.
K-means clustering technique is then used to reduce the number of scenarios.}

The single-stage scenario set for stage $t$+1 that is developed from a certain node at stage $t$ is different from the ones developed from the other nodes at stage $t$,
since the past trend of uncertain parameters affect their future behavior.
Finally, each multi-stage scenario (hereinafter called scenario) is a full trace of uncertain parameters
from one of the end nodes at stage $t$=$\tau$+$T$ back to the starting node at stage $t$=$\tau$. Considering the
effects of the parent node in the single stage scenario generation process reduces the chance of unlikely
variations of uncertain parameters between successive periods. This is a matter of premium importance when scheduling
the \textsc{dcd}s, since the number of \textsc{dca}s and therefore the cost of these actions depend on the rate of variation in uncertain parameters.
\vspace{-2.2mm}
\subsection{Solution Methodology}\label{Sol}
The main focus of this paper is on providing a strategic framework for co-scheduling the fast and discrete control devices. To keep the narrative simple, some muddling aspects are neglected bearing in mind that the proposed methodology should be able to solve the \textsc{dsp} with all these aspects included. It is assumed that the system is balanced. The details of applying \textsc{tra} for solving unbalanced \textsc{dsp} were provided in \cite{Sheng2014}.
The discrete variables include the tap positions, steps of \textsc{cb}s and additional binary variables introduced to model the tap-/step-changing costs. If the originally linear models of \textsc{dcd}s (subsection \ref{Slow_Control}) are applied, the respective auxiliary binary variables are also included in the list of discrete variables.

\begin{figure}[!t]
  \centering
  \includegraphics[width=0.90\columnwidth, trim=0.00cm 0.15cm 0.0cm 0.15cm, clip=true]{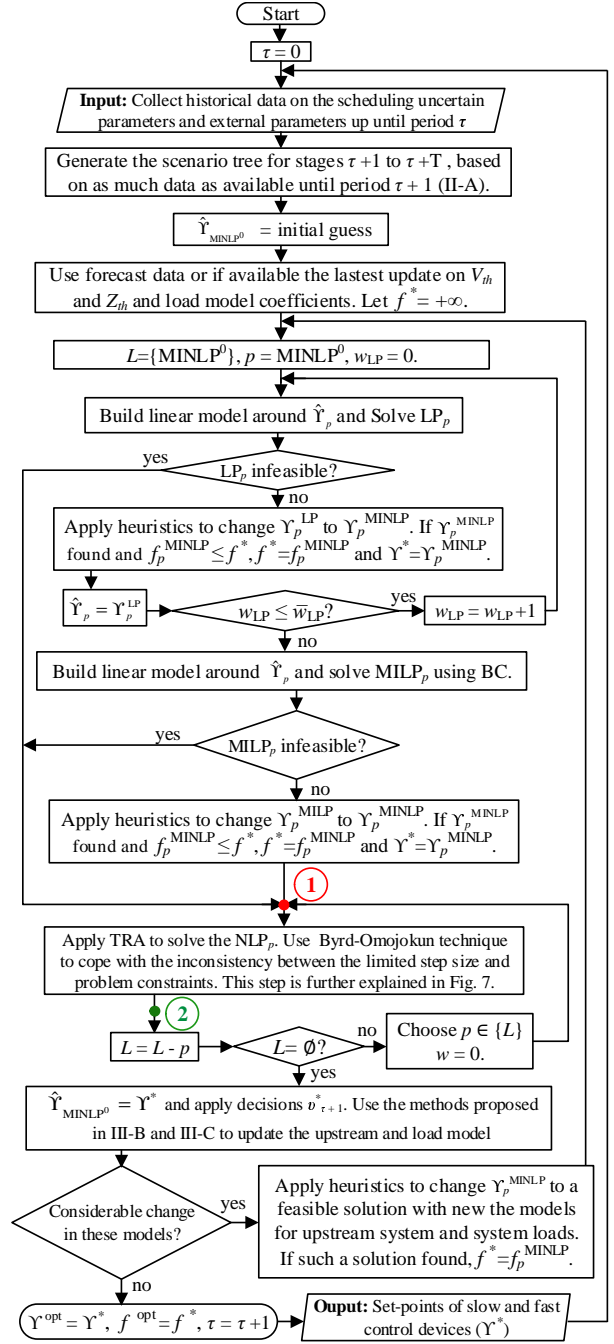}
  \vspace{-1mm}
  \caption{\textcolor[rgb]{0,0,0}{Comprehensive flowchart of the proposed scheduling algorithm.}}
	\vspace{-3mm}
	\label{Flow1}
\end{figure}

\textcolor[rgb]{0,0,0}{The proposed scheduling algorithm is explained step by step in the upcoming sections.
An overview of this algorithm is provided here based on the comprehensive flowchart of Fig. \ref{Flow1}.
\textsc{bc} technique is applied to simultaneously handle the integer and continuous variables.
In Fig. \ref{Flow1}, set $L$ (indexed by $p$) is the set of all \textsc{minlp} problems that should be solved during the branching process. The original problem (at the root node of \textsc{bc}) is given be $\text{MINLP}^0$. $\Upsilon^*$ is the best integer feasible solution found so far during the branching process and $f^*=f(\Upsilon^*)$. In the final optimal solution point, $f=f^\text{opt}$ and $\Upsilon=\Upsilon^\text{opt}$. The optimal solution of \textsc{nlp} problem $p$, optimal solution of \textsc{lp} problem $p$ and a feasible solution for \textsc{minlp} problem $p$ are given by $\Upsilon_p^\text{NLP}$, $\Upsilon_p^\text{LP}$ and $\Upsilon_p^\text{MINLP}$, respectively.}

\textcolor[rgb]{0,0,0}{In Fig. \ref{Flow1}, the input data for the scheduling window $\tau < t \leq \tau+T$ includes the historical data on all uncertain parameters that affect the future of the scheduling problem uncertain parameters. The input data can be divided into two categories. The first category, i.e., the historical data on the scheduling problem uncertain parameters, includes the active and reactive demands, load model parameters, utility power purchase prices, upstream system model parameters and production of each \textsc{res} up until period $\tau$. For the upcoming periods $t$ ($\tau+1 \leq t \leq \tau+T$) these parameters are given in the scenario vectors. The second category of the input parameters includes the historical data on the external parameters. The previous values of these parameters may affect the forecast of the scheduling problem uncertain parameters for the upcoming periods. For instance, the wind speed and solar radiation in the neighbor areas during the periods up until $\tau$ affect the forecast of wind and solar power generations within the intended system.}

\textcolor[rgb]{0,0,0}{In Fig. \ref{Flow1}, after generating the scenarios (see subsection \ref{Uncertain}), a few steps are first taken to expedite the solution. The aim is to find a high quality initial solution and a tight upper bound on the value of objective function of the \textsc{minlp} problem. The tighter this bound, the lower the branching burden. 
A simplified (linearized) problem is first solved using \textsc{lp} technique.
A heuristic approach is then applied to change the solution to a feasible \textsc{minlp} solution.
In this approach, the values of discrete variables are first rounded-off to the nearest integer values.
\textsc{tra} is next applied to solve the resultant \textsc{nlp} problem and to find the values of continuous variables.
Under such setup, sometimes a tighter upper bound is found for the \textsc{minlp} objective function.
This step is not outlined in Fig. \ref{Flow1} and is referred to as ``applying heuristics to change $\Upsilon_p^\text{LP}$ to $\Upsilon_p^\text{MINLP}$''.}

\textcolor[rgb]{0,0,0}{If this step finds $\Upsilon_p^\text{MINLP}$ and $f^\text{MINLP}<f^*$, $f^*$ is replaced with $f^\text{MINLP}$.
This procedure is repeated a few times to increase the chance of obtaining a tighter upper bound. The solution found in the previous \textsc{lp} is considered as the initial solution of the next one.
The counter $w_\text{LP}$ is used to count the \textsc{lp}s solved. The maximum number of \textsc{lp}s that are solved at this step is $\overline{w}_\text{LP}$.
In the case studies, $\overline{w}_\text{LP}$=2. This $w_\text{LP}$ is selected based on the experience gained by solving the scheduling problems in the studies of Section \ref{Case_Studies}.
With higher $\overline{w}_\text{LP}$, the chance of finding a tighter bound increases. However, the higher number of \textsc{lp}s that should be solved may increase the solution time.
Moreover, it was observed that with more than three attempts ($\overline{w}_\text{LP}>2$), sometimes, the solution of the \textsc{lp} problem starts to fluctuate.
The value of $\overline{w}_\text{LP}$ is set to two, so as to allow the algorithm a fair chance of attaining a tight upper bound while avoiding fluctuation and high solution times.}

\begin{figure}[!t]
  \centering
  \includegraphics[width=0.86\columnwidth, trim=0.00cm 0.11cm 0.0cm 0.07cm, clip=true]{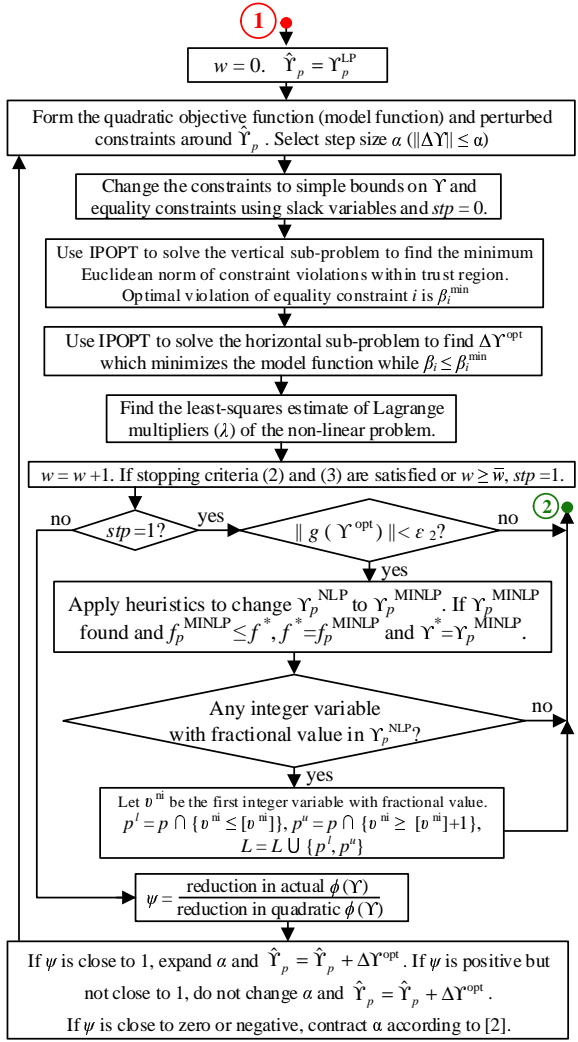}
  \vspace{-1mm}
  \caption{\textcolor[rgb]{0,0,0}{Steps of \textsc{tra} for solving \textsc{nlp} problems.}}
	\vspace{-7.5mm}
	\label{Flow3}
\end{figure}

\textcolor[rgb]{0,0,0}{It is discussed in Section \ref{Deter_Alg} how to find the model function (quadratic objective function) and perturbed objective function for \textsc{tra} and simplified problem, respectively.
A \textsc{milp} problem is also solved using \textsc{bc} to further increase the chance of attaining a tight bound. This explains the steps of Fig. \ref{Flow3} up to point \circled{1}. The application of \textsc{tra} to solve the \textsc{nlp}s during the branching process of \textsc{bc} is next explained. The respective steps are summarized in Fig. \ref{Flow1} between points \circled{1} and \circled{2}. These steps are better described in Fig. \ref{Flow3}.}

\textcolor[rgb]{0,0,0}{In Fig. \ref{Flow3}, a globally convergent \textsc{tra} is applied to solve the integer relaxed \textsc{nlp} problem at each node.
The objective function \eqref{nlp} includes the operation cost, control cost and penalty terms associated with voltage deviations.
The voltage deviation constraints are modeled as soft constraints using these penalty terms.
The quadratic model function and perturbed constraints are built for each \textsc{tra} sub-problem based on Section \ref{Deter_Alg}.
\textsc{tra} first changes the inequality constraints to equality constraints \cite{BAHRAMI2020}, i.e., $g(\Upsilon)=0$ in \eqref{nlp}, using auxiliary variables. The simple bounds on optimization variables, i.e., $\underline{\Upsilon}$$\leq$$\Upsilon$$\leq$$\overline{\Upsilon}$ can be directly handled.
In each sub-problem, \textsc{tra} minimizes a model function, subjected to the perturbed constraints, within a trust region around the candidate solution of the previous sub-problem.}

\textcolor[rgb]{0,0,0}{In Fig. \ref{Flow3}, $\left\|G\right\|$ gives the Euclidean norm of vector $G$. The integer part of real variable $\upsilon$ is given by $[\upsilon]$.
In each sub-problem of \textsc{tra}, the problem constraints might be inconsistent with the step-size ($||\Delta \Psi|| \leq \alpha$).
Byrd-Omojokun technique \cite{BAHRAMI2020} is used to cope with the inconsistency.
To this end, each sub-problem is divided into the vertical and horizontal sub-problems in Fig. \ref{Flow3}.
In the vertical sub-problem, the objective is to minimize the Euclidean norm of constraint violations within the trust region.
The result of this step includes the optimal constraint violation $\beta_i^\text{min}$ for each constraint $g_i(\Upsilon)=0$.}

\begin{equation}
\label{nlp}
\begin{array}{l}
\displaystyle
\hspace{-3.5mm}
\underbrace{Min}_{\Upsilon}
\hspace{0.5mm}
f(\Upsilon) \hspace{-1mm} = \hspace{-1mm} \sum_{s=1}^{N_s} \hspace{-0.5mm} \pi_s \hspace{-1.25mm}  \sum_{t=\tau+1}^{T+\tau} \hspace{-1mm} [ \sigma_t OC_{t,s} \hspace{-0.75mm}
+ \hspace{-0.75mm} CC_{t,s} \hspace{-0.75mm} + \hspace{-1mm} \sum_{b=1}^{N_b} \hspace{-0.25mm} \eta_b {VD_b}_{t,s} ] \\
\displaystyle
\hspace{7mm} \textcolor[rgb]{0,0,0}{\text{s.t.:} \hspace{5mm} g(\Upsilon)=0,} 
\hspace{7mm} \textcolor[rgb]{0,0,0}{\underline{\Upsilon} \leq \Upsilon \leq \overline{\Upsilon} }
\\
\displaystyle
\textcolor[rgb]{0,0,0}{\Upsilon=\left\{\upsilon_{\tau+1},\hspace{2mm} \upsilon_{\tau+1,s},\hspace{2mm} \upsilon_{t,s} \hspace{3mm} {\forall \ \tau+2 \leq t \leq T+\tau}\right\}}
\end{array}
\hspace{-2mm}
\end{equation}
\begin{equation}
\label{StopCr1}
\begin{array}{l}
\displaystyle
\textcolor[rgb]{0,0,0}{  \text{stopping criterion 1:} \hspace{3mm} \left\| \nabla f(\Upsilon)+ \nabla g(\Upsilon)^T \lambda \right\|<\epsilon_1}
\end{array}
\end{equation}
\begin{equation}
\label{StopCr2}
\begin{array}{l}
\displaystyle
\hspace{-12mm}
\textcolor[rgb]{0,0,0}{ \text{stopping criterion 2:} \hspace{17mm} \left\| g(\Upsilon) \right\|<\epsilon_2}
\end{array}
\end{equation}
\begin{equation}
\label{mrt}
\begin{array}{l}
\displaystyle
\textcolor[rgb]{0,0,0}{\phi(\Upsilon)=f(\Upsilon) + \zeta \left\| g(\Upsilon) \right\|}
\end{array}
\end{equation}
\vspace{-5mm}

\textcolor[rgb]{0,0,0}{The number of \textsc{tra} sub-problems solved in order to solve this \textsc{nlp} problem, is given by $w$. The maximum number of iterations is $\overline{w}$. 
The Lagrange optimality and constraints' satisfaction conditions are provided in \eqref{StopCr1} and \eqref{StopCr2}, respectively.
After each sub-problem, if \eqref{StopCr1} and \eqref{StopCr2} are simultaneously satisfied or $w \geq \overline{w}$, \textsc{tra} is stopped.
The Lagrange multipliers of the equality constraints of \eqref{nlp} are given by vector $\lambda$.
These multipliers are not computed by \textsc{tra}. A least-squares estimate is used to find $\lambda$ based on \cite{BAHRAMI2020}.
To decide on the step size for the next \textsc{tra} sub-problem, parameter $\psi$ is used according to Fig. \ref{Flow3} and \cite{BAHRAMI2020}.
The merit function $\phi(\Upsilon)$ is provided in \eqref{mrt}. $\zeta \geq 1$ is a penalty parameter that weights constraint satisfaction against objective minimization.}

\textcolor[rgb]{0,0,0}{After finding the optimal solution of the \textsc{minlp} problem, the parameters of upstream system and load models are updated by comparing the measured voltages and currents before and after applying $\upsilon_{\tau+1}$ (see subsections \ref{Slow_Control} and \ref{Load_MDL}).
If the changes of these models are higher than the predefined tolerances, the \textsc{minlp} problem is solved again considering the previous solution as the initial guess to give \textsc{bc} a warm start.
Only the decisions made for period $\tau+1$ are applied and the operator waits for further information to make the next decisions.}

\textcolor[rgb]{0,0,0}{The expediting \textsc{lp}s are solved using CPLEX under GAMS.
The vertical and horizontal sub-problems of \textsc{tra} are solved using IPOPT under GAMS.
The scheduling algorithm is implemented in MATLAB on a PC with an Intel(R) Xeon(R) E5-1650 3.6 GHz CPU and 16 GB of RAM.}
\section{Deterministic Scheduling}\label{Deter_Alg} 
The fast control devices and \textsc{cb}s are modeled as controllable current sources.
\textsc{oltc} transformer and upstream system are modeled together as a controllable current source. The network is modeled by admittance matrix.
The system loads are modeled as dependent current sources. The load and upstream system models are kept updated while solving the \textsc{dsp}.
\subsection{Fast Control Devices}\label{Fast_Control}
For fast control devices, the independent control variables include $P_g$ and $Q_g$.
The perturbed model is presented in \eqref{DER_Linear}-\eqref{A_SVCDERRR} for a dispatchable \textsc{der}.
$I$ and $V$ are the current injected by this \textsc{der} and the voltage at the connecting bus.
Subscripts $x$ and $y$ give the real and imaginary parts, respectively.
\textcolor[rgb]{0,0,0}{\textsc{svr}s cannot exchange active power with network. It is also assumed that \textsc{res}s produce as much power as the respective natural sources and their capacity constraints allow for. Thus, for \textsc{svr}s and \textsc{res}s} the only controllable parameter is $Q_g$
and the second term in the right hand side of \eqref{DER_Linear} should be replaced with the first column of matrix $B$ multiplied by $\Delta Q_g$.
\begin{equation}
\label{DER_Linear}
\begin{array}{l}
\displaystyle
\hspace{-3mm}
\begin{pmatrix} \Delta{I_x} \\ \Delta{I_y}\end{pmatrix} \hspace{-1mm}
= \hspace{-1mm}
A_{2\times{2}}  \hspace{-1mm}
\begin{pmatrix} \Delta{V_x} \\ \Delta{V_y} \end{pmatrix} \hspace{-1mm}
+\hspace{-1mm}
B_{2\times{2}}  \hspace{-1mm}
\begin{pmatrix} \Delta{P}_g \\ \Delta{Q}_g \end{pmatrix}
\end{array}
\hspace{-1mm}
\end{equation}
\vspace{-2mm}
\begin{equation}
\label{A_SVCDERRR}
\begin{array}{l}
\displaystyle
\hspace{-3mm} A \hspace{-1mm}
= \hspace{-1mm}
- \hspace{-1mm}
\begin{pmatrix} \hat{V}_x & \hat{V}_y\\ \hat{V}_y & -\hat{V}_x \end{pmatrix}^{-1} 
 \hspace{-1mm}
\begin{pmatrix} \hat{I}_x & \hat{I}_y\\ -\hat{I}_y & \hat{I}_x \end{pmatrix},
\hspace{3mm}
B  \hspace{-1mm}
= \hspace{-1mm}
\begin{pmatrix} \hat{V}_x \hspace{-2mm} & \hspace{-2mm} \hat{V}_y \\ \hat{V}_y \hspace{-2mm} & \hspace{-2mm} -\hat{V}_x \end{pmatrix}^{-1}
\end{array}
\end{equation}

\begin{figure}[!t]
  \centering
  \includegraphics[width=0.58\columnwidth, angle =-90, clip=true]{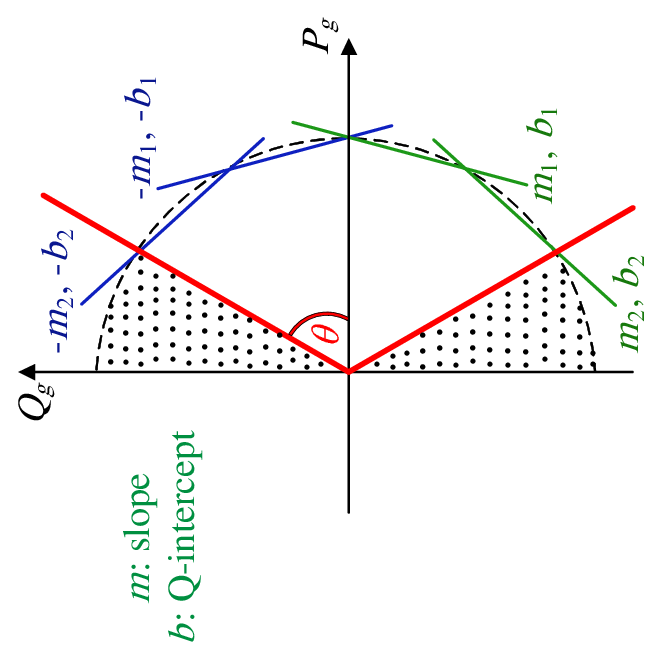} \vspace{-3mm}
  \caption{Set of linear capacity constraints on dispatchable \textsc{der}s with $n$=1,2. In the case studies, the number of linear constraints is 10, i.e., $\overline{n}=5$.}\label{DisDER}  
	\vspace{-3mm}
\end{figure}

High quality perturbed formulations should also be developed for capacity constraints. The active power production of each \textsc{res} is given in scenario vectors. The apparent power limitation of \textsc{res}s is presented in \eqref{RESLimit}. For each PV unit, considering the maximum power angle of $\overline{\alpha}^{pv}$ 
constraints \eqref{PV_PF} should be held \textcolor[rgb]{0,0,0}{to avoid high harmonic distortions.} \textcolor[rgb]{0,0,0}{As shown in \eqref{DFIG},} for a doubly-fed induction wind generator, the reactive power injection cannot be lower than a specified value.
For dispatchable \textsc{der}s, linear representation \eqref{Cap_Limit} is extracted with a desired level of accuracy using the model presented in Fig. \ref{DisDER}.
The dash-lined semicircle models the maximum apparent power that this \textsc{der} can provide. This constraint is approximated by desired number of lines ($m_n$, $b_n$), where $m_n$ and $b_n$ are the slope and $Q$-intercept of line $n$. These constraints are linear and therefore, the regarding perturbed constraints are accurate. \textcolor[rgb]{0,0,0}{Based on Fig. \ref{DisDER}, if the maximum apparent power constraint of \textsc{der} $e$ is approximated by $2\hspace{0.5mm}\overline{n}$ linear constraints, the maximum error of such approximation is $(1-cos(\theta/4\overline{n}))\overline{S}^e$. For instance, with $\theta=36.87^\text{o}$, i.e., minimum power factor of 0.8, the maximum errors are $0.0032\overline{S}^e$ and $0.00051\overline{S}^e$ with $\overline{n}$=2 and $\overline{n}$=5, respectively. As can be seen, this linear representation provides a quite acceptable approximation. The level of accuracy demanded by the decision maker determines $\overline{n}$ based on the relationship provided for the maximum error.}
\begin{equation}
\label{RESLimit}
\displaystyle
\pm \left(\hat{Q}_g^{res}+\Delta Q_g^{res} \right)
\leq 
\sqrt{
\left(\overline{S}^{res}\right)^2 
- \hspace{-1mm}
\left(P_g^{res}\right)^2} 
\end{equation}
\begin{equation}
\label{PV_PF}
\displaystyle
-tan(\overline{\alpha}^{pv}) P_g^{pv}
\leq
\hat{Q}_g^{pv}+
\Delta Q_g^{pv} \hspace{-1.5mm}
\leq
tan(\overline{\alpha}^{pv}) P_g^{pv}
\end{equation}
\begin{equation}
\label{DFIG}
\displaystyle
\underline{Q}^{wind}
\leq
\hat{Q}_g^{wind}+
\Delta Q_g^{wind} \hspace{-1.5mm}
\end{equation}
\begin{equation}
\label{Cap_Limit}
\begin{array}{l}
\displaystyle
\hspace{0mm}
Q_g^{der} \leq tan(\theta)P_g^{der}, \hspace{5mm} Q_g^{der} \geq -tan(\theta)P_g^{der}\\
\displaystyle
\hspace{-3mm}
Q_g^{der} \geq m_n P_g^{der}+ b_n, \hspace{3.5mm} Q_g^{der} \leq -m_n P_g^{der}- b_n  \hspace{3mm} \forall n
\end{array}
\end{equation}
\vspace{-3mm}
\begin{figure}[!t]
  \centering
  \includegraphics[width=0.6\columnwidth,trim=0.0cm 0.05cm 0.0cm 0.20cm,clip=true]{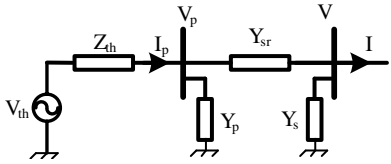} \vspace{-3mm}
  \caption{Upstream network and \textsc{oltc} transformer $pi$ models.}\label{Up_OLTC}
	\vspace{-3.5mm}
\end{figure}
\subsection{Slow Control Devices}\label{Slow_Control}
\textbf{\textsc{oltc} transformer model 1}: \textsc{oltc} control affects the transformer model. A perturbed model is developed here for \textsc{oltc} transformers. Fig. \ref{Up_OLTC} shows the $pi$ model of an \textsc{oltc} transformer connected to an upstream system with Thevenin voltage and impedance of $V_{th}$ and $Z_{th}$, respectively. A $pi$ model was presented in \cite{Nouri2019} for \textsc{oltc} transformers. This model is extended to include the magnetizing current and core loss. The admittances of the model presented in Fig. \ref{Up_OLTC} is given in \eqref{OLTC_Eq}.

\textcolor[rgb]{0,0,0}{It is assumed that the tap changer has been installed on the primary winding, i.e., the number of secondary turns is fixed. The transformer core is assumed to remain unsaturated. As the tap position ($tap$) increases, the number of primary turns and the turn ratio ($r$) increase. In the nominal tap position (given by superscript $n$), $tap^n$=0 and the turn ratio (in pu.) is 1 ($r^n$=1).
Tap changing operations change the per unit value of the primary series impedance ($Z_{sr,p}$) proportional to the number of primary turns. The per unit value of the secondary series impedance ($Z_{sr,s}$), $X_M$ and $R_c$ viewed from the secondary terminal do not change.
It is assumed that for a well manufactured transformer $Z_{sr,p}^n=Z_{sr,s}^n=Z_{sr}^n$/2 \cite{Nouri2019}.}

\textcolor[rgb]{0,0,0}{For \textsc{oltc} transformers, the independent optimization variable is $tap$ which is an integer variable and $tap=(r-1)/{\Delta}U$.}
For parallel transformers, each admittance in the $pi$ model is equal to the sum of regarding admittances for all transformers.
\vspace{-1mm}
\begin{equation}
\label{OLTC_Eq}
\displaystyle
\hspace{-0.0mm}
Y_{sr} \hspace{-1mm} = \hspace{-1mm} \frac{1}{rZ_{sr}^n}, \hspace{2.25mm}
Y_p \hspace{-1mm} = \hspace{-1mm} \frac{(1 \hspace{-0.75mm} - \hspace{-0.5mm} r)}{r^2Z_{sr}^n} \hspace{-0.5mm} + \hspace{-0.5mm} \frac{1}{r^2R_c^n} \hspace{-0.5mm} - \hspace{-0.25mm} {\frac{\text{j}}{r^2X_M^n}}, \hspace{2.25mm}
Y_s \hspace{-1mm} = \hspace{-1mm} {\frac{(r \hspace{-0.5mm} - \hspace{-0.5mm} 1)}{rZ_{sr}^n}} \hspace{-1.75mm}
\end{equation}
\begin{equation}
\label{OLTC_Control}
\displaystyle
I=C(tap)V+D(tap)V_{th}
\end{equation}
\begin{equation}
\label{Linear_OLTC}
\displaystyle
\begin{pmatrix} \Delta{I_x} \\ \Delta{I_y} \end{pmatrix} \hspace{-1mm} = \hspace{-1mm}
A_{2\times{2}} \hspace{-1mm} \begin{pmatrix} \Delta{V_x} \\ \Delta{V_y} \end{pmatrix} \hspace{-1mm}
+ \hspace{-1mm}
B_{2\times{1}} \Delta{tap}, \hspace{3mm}
A \hspace{-1mm}= \hspace{-1.5mm}
\begin{pmatrix} \hspace{-0.5mm} C_x \hspace{-1.5mm} & \hspace{-2mm} -C_y \hspace{-2mm} \\ \hspace{-0.5mm} C_y \hspace{-1.5mm} & \hspace{-2mm} C_x \hspace{-2mm}
\end{pmatrix}\hspace{-1mm}
\end{equation}
\begin{equation}
\label{Linear_OLTC_AB}
\displaystyle
B \hspace{-0.5mm}
= \hspace{-1.0mm}
\begin{pmatrix}
V_x\frac{\partial{C_x}}{\partial{tap}} \hspace{-0.25mm}
+ \hspace{-0.25mm}
{V_{th}}_x\frac{\partial{D_x}}{\partial{tap}} \hspace{-0.25mm}
- \hspace{-0.25mm}
V_y.\frac{\partial{C_y}}{\partial{tap}} \hspace{-0.5mm}
- \hspace{-0.25mm}
{V_{th}}_y\frac{\partial{D_y}}{\partial{tap}} \hspace{-0.25mm} \vspace{1mm}
\\
V_y\frac{\partial{C_x}}{\partial{tap}} \hspace{-0.25mm}
+ \hspace{-0.25mm}
{V_{th}}_y\frac{\partial{D_x}}{\partial{tap}} \hspace{-0.25mm}
+ \hspace{-0.25mm}
V_x\frac{\partial{C_y}}{\partial{yap}} \hspace{-0.25mm}
+ \hspace{-0.25mm}
{V_{th}}_x\frac{\partial{D_y}}{\partial{tap}} \hspace{-1mm} 
\end{pmatrix}
\end{equation}

A perturbed relationship between $I$, $V$ and $tap$ is given in \eqref{Linear_OLTC}. Matrices $A$ and $B$ are defined in \eqref{Linear_OLTC} and \eqref{Linear_OLTC_AB}, respectively.
The relationship between the voltage and current at the secondary bus and the \textsc{oltc} control variables ($tap$) is shown in \eqref{OLTC_Control}, where $C$ and $D$ can be found using Fig. \ref{Up_OLTC}.

In the proposed framework, $V_{th}$ and  $Z_{th}$ are given in scenarios for $t=\tau+2$:$\tau+T$. For $t=\tau+1$, it is necessary to find the accurate values of these parameters, since they affect the system model.
\textcolor[rgb]{0,0,0}{It is important to keep $V_{th}$ and $Z_{th}$ up to date, while solving the \textsc{dsp}.
By tracking the variations of measured (or estimated) $V_p$ and $I_p$, $V_{th}$ and $Z_{th}$ can be found, if the variations of $V_p$ and $I_p$ are caused dominantly by a change in the downstream network \cite{Bahadornejad2014}. However, during the normal operation, $V_p$ and $I_p$ gradually change 
and it is not possible to understand if the source of these changes is in the downstream or upstream systems.}

\textcolor[rgb]{0,0,0}{Based on \cite{Bahadornejad2014}, $Z_{th}$ can be found by changing the control variables in downstream network and measuring (estimating) $V_p$ and $I_p$ before and after applying this changes.
The changes should be significant enough to cancel the effects of measurement errors.
The measurement instants should also be as close as possible to not allow the upstream system changes to affect the measured data.
Here, both $V_{th}$ and $Z_{th}$ are found by comparing the measured values of $V_p$ and $I_p$, before and after applying the changes proposed by the scheduling algorithm.}

\textcolor[rgb]{0,0,0}{Equation \eqref{Thevinin_model_b} gives the relationship between $V_{th}$, $Z_{th}$, $V_p$ and $I_p$. Superscripts $\varsigma$ takes the values 0, 1, and 2. For the values before applying the changes proposed by the scheduling algorithm, $\varsigma$=0. For the values after applying the first and second changes, $\varsigma$ is 1 and 2, respectively.
In \eqref{Thevinin_model_b}, the measured values are distinguished by a bar upon them. There are six variables, i.e., $\delta_{Z_{th}}$, $\delta_{V_{p}^0}$, $\delta_{V_{p}^1}$, $\delta_{V_{p}^2}$, $V_{th}$ and $Z_{th}$. With the values of 0, 1, and 2 for $\varsigma$, \eqref{Thevinin_model_b} gives three equations. They are rewritten in six equations separating the real and imaginary parts. Thus, $V_{th}$ and $Z_{th}$ are found.
If the changes in $V_{th}$ and $Z_{th}$ are higher than a predefined level, the proposed method is applied again to solve the \textsc{dsp} with updated upstream model (see \ref{Sol}).} 
\begin{equation}
\label{Thevinin_model_b}
\displaystyle
\left|V_{th}\right|^{\angle{0}}=\overline{\left|V_{p}^\varsigma \right|}^{\angle{\delta_{V_{p}^\varsigma}}}   +Z_{th}^{\angle{\delta_{Z_{th}}}}.\overline{\left|I_{p}^\varsigma \right|}^{\angle{\delta_{V_{p}^\varsigma}}+\overline{\varphi_{I_{p}^\varsigma,V_{p}^\varsigma}}}
\end{equation}

\textbf{\textsc{oltc} transformer model 2}: An originally linear representation is developed here for the governing relationships
of \textsc{oltc} transformers using auxiliary binary variables. The results of applying this model and those obtained using
the perturbed model already developed are compared in the case studies.

\begin{figure}[!t]
  \centering
  \includegraphics[width=0.96\columnwidth, trim=0.0cm 0.25cm 0.0cm 0.10cm, clip=true]{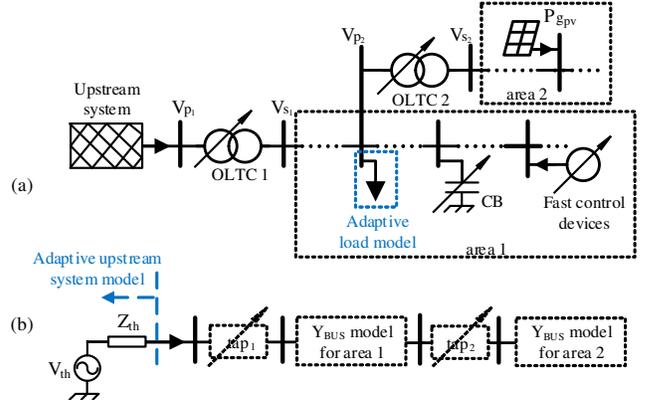}
	\vspace{-2mm}
  \caption{a) A simple active distribution system, b) how to model \textsc{oltc} transformers in active distribution systems.}\label{TransYbus0}
	\vspace{-3mm}
\end{figure}
Both this model and the previous perturbed model can be used for the \textsc{oltc} transformers like transformers 1 and 2 in
Fig. \ref{TransYbus0}(a) for which both primary and secondary voltages are variable,
i.e., depend on control variables. For instance, ${V_p}_1$ depends on the voltage drop across $Z_{th}$
and in turn depends on all control variables. Fig. \ref{TransYbus0}(b) shows how the transformer models
should be combined with the $\text{Y}_\text{BUS}$ representations of the other parts of system and
also the adaptive upstream model.

\begin{figure}[t!]
  \centering
\includegraphics[clip = true,width=0.65\columnwidth, trim=0.0cm 0.03cm 0.0cm 0.23cm,
]{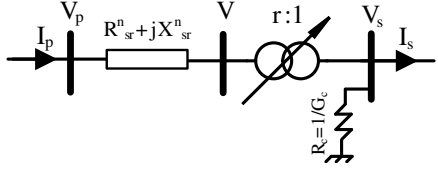}
\vspace{-2.5mm}
  \caption{\textsc{oltc} transformer model 2.}
\vspace{-3.75mm}
\label{TransM}
\end{figure}

Fig. \ref{TransM}, presents \textsc{oltc} transformer model used to extract the originally linear transformer equations.
$X_M$ and also the effects of tap position on series impedance are neglected.
\textcolor[rgb]{0,0,0}{Core resistance $R_c$ and the effects of tap-changing operations on core loss are included.}
This model is formulated in \eqref{tr1}. The terms $rV_s$ and $rI_p$ are nonlinear which are linearized here without loss of accuracy.
For \textcolor[rgb]{0,0,0}{brevity}, the formulation is just presented for imaginary components.
\textcolor[rgb]{0,0,0}{For real components, the same formulation can be extracted.}

Variable $r$ is a non-integer discrete variable which is rewritten in terms of dummy binary variables $\nu_m$ in \eqref{tr2}.
The minimum turn ratio is $\underline{r}$. 
The linear counterpart of first equation in \eqref{tr1} is provided in \eqref{tr3} (imaginary component).
Linear terms $\gamma_m$ satisfies \eqref{tr5}, where $\sigma$ is a binary variable. For ${V_s}_y \geq 0$, $\sigma$=1. $\overline{{V_s}_y}$ is an upper bound for ${V_s}_y$.
Non-linearitiy of $rI_p$ is similarly dealt with.
This model is applied separately for each phase
for independent per-phase \textsc{oltc}s in unbalanced systems.

\begin{equation}
\label{tr1}
\begin{array}{l}
\displaystyle

\hspace{-3.5mm}
V \hspace{-0.5mm} = \hspace{-0.5mm} rV_s, \hspace{3.5mm}
I_s \hspace{-0.5mm} = \hspace{-0.5mm} rI_p \hspace{-0.5mm} + \hspace{-0.5mm} G_cV_s, \hspace{3.5mm}
V \hspace{-0.5mm} = \hspace{-0.5mm} V_p \hspace{-0.5mm} + \hspace{-0.5mm}
\left(R_s \hspace{-0.5mm} + \hspace{-0.5mm} jX_s\right)I_p

\end{array}
\hspace{-3mm}
\end{equation}
\vspace{-2mm}
\begin{equation}
\label{tr2}
\begin{array}{l}
\displaystyle
\hspace{-3mm}
r=\underbrace{1-N_m^- \Delta U}_{\underline{r}} +\left(\sum_{m=1}^{N_m} \nu_m\right)\Delta U, \hspace{3.5mm}
\nu_m \leq \nu_{m-1} \hspace{1.5mm} \forall m

\end{array}
\hspace{-3mm}
\end{equation}
\vspace{-3mm}
\begin{equation}
\label{tr3}
\begin{array}{l}
\displaystyle
\hspace{-3mm}

{V}_y=\underline{r}{V_s}_y+\Delta U \sum_{m=1}^{N_m} \gamma_m

\end{array}
\hspace{-3mm}
\end{equation}
\vspace{-2mm}
\begin{equation}
\label{tr5}
\begin{array}{l}
\displaystyle
\hspace{-5mm}
(\sigma-1)\overline{{V_s}_y}\leq {V_s}_y \leq \sigma \overline{{V_s}_y}
\vspace{1mm}\\
\displaystyle
\hspace{-5mm}
-\nu_m \overline{{V_s}_y} \leq \gamma_m \leq \nu_m \overline{{V_s}_y}
\vspace{1mm}\\
\displaystyle
\hspace{-5mm}
\sigma \overline{{V_s}_y}+{V_s}_y \leq \gamma_m \leq (1-\sigma) \overline{{V_s}_y}+{V_s}_y 
\vspace{1mm}\\
\displaystyle
\hspace{-5mm}
(\sigma-1) \overline{{V_s}_y} \leq \gamma_m \leq \sigma \overline{{V_s}_y}
\vspace{1mm}\\
\displaystyle
\hspace{-5mm}
{V_s}_y \hspace{-0.5mm} + \hspace{-0.5mm} \overline{{V_s}_y}\nu_m \hspace{-0.5mm} - \hspace{-0.5mm} \overline{{V_s}_y}\sigma \hspace{-0.5mm}
\leq \gamma_m \leq
{V_s}_y \hspace{-0.5mm} - \hspace{-0.5mm} \overline{{V_s}_y}\nu_m \hspace{-0.5mm} + \hspace{-0.5mm} \overline{{V_s}_y}\sigma \hspace{-0.5mm} + \hspace{-0.5mm} \overline{{V_s}_y}
\end{array}
\hspace{-3mm}
\end{equation}
\vspace{-1mm}

\textbf{Perturbed and originally linear models for \textsc{cb}s}: The reactive power of \textsc{cb}s is a function of their impedances and voltages.
The perturbed model of each \textsc{cb} is provided in \eqref{CBs_Linear} and \eqref{CBs_AB}.
An originally linear model is also developed for \textsc{CB}s for the sake of comparison. The non-linear model is $I^{cb}$=$\text{j}y_{st}.stV^{cb}$. \textcolor[rgb]{0,0,0}{The term $stV^{cb}$ is the product of integer variable $st$ and continuous variable $V^{cb}$.} The process of extracting the originally linear model is the same as the one used for the first equation of \eqref{tr1}.
\begin{equation}
\label{CBs_Linear}
\begin{array}{l}
\displaystyle
\hspace{-3mm}
\begin{pmatrix} \Delta{I_x} \\ \Delta{I_y}\end{pmatrix} \hspace{-1mm}
= \hspace{-1mm}
A^{cb}_{2\times 2}
\begin{pmatrix} \Delta{V_x} \\ \Delta{V_y} \end{pmatrix} \hspace{-1mm}
+\hspace{-1mm}
B^{cb}_{2\times 1}
\Delta{st}
\end{array}
\hspace{-1mm}
\end{equation}
\vspace{-2mm}
\begin{equation}
\label{CBs_AB}
\begin{array}{l}
\displaystyle
\hspace{-3mm}
A^{cb}\hspace{-1mm}
=
\begin{pmatrix} 0 & st.y^{st} \\ -st.y^{st} & 0 \end{pmatrix}, \hspace{7mm}
B^{cb}\hspace{-1mm}
=
\begin{pmatrix} y^{st}V_y \\ -y^{st}V_x \end{pmatrix} \hspace{-1mm}
\end{array}
\hspace{-1mm}
\end{equation}
\vspace{-6mm}
\subsection{Adaptive Load Model}\label{Load_MDL}
Power demands depend on the voltage levels at load points. Load model accuracy affects the solution optimality.
With an inaccurate model, the expected energy saving is not realized.
ZIP model \eqref{ZIP} which decomposes the system loads into constant impedance, constant current and constant power components \cite{Nouri2017} is applied here. In \eqref{ZIP}, $a_p+b_p+c_p=a_q+b_q+c_q=1$.
\textcolor[rgb]{0,0,0}{To keep the simplicity of presentation, this ZIP load model is replaced by a ZP model. Replacing $\left|V\right|$/$V_0$ by 0.5(1+$\left|V\right|^2$/$V_0^2)$ in \eqref{ZIP}, ZP load model is provided in \eqref{ZP}. Comparing \eqref{ZP} with \eqref{ZIP}, $a'_p$=$a_p$+$b_p$/2, $c'_p$=$c_p$+$b_p$/2 and $a'_p+c'_p=a'_q+c'_q=1$. Within the typical range of bus voltages in the steady state studies, the accuracies of ZIP and ZP models are quite close and \eqref{ZP} gives a quite acceptable approximation of \eqref{ZIP}.
For instance, with $\left|V\right|$/$V_0$=0.9 which is a voltage drop condition that might never happen in the steady state operation of distribution systems, 0.5(1+$\left|V\right|^2$/$V_0^2)$=0.905.
For the voltages closer to unity, this approximation is even more accurate.}

\begin{figure}[!t]
  \centering
  \includegraphics*[width=0.75\columnwidth,trim=0.05cm 0.05cm 0.5cm 0.3cm,clip=true]{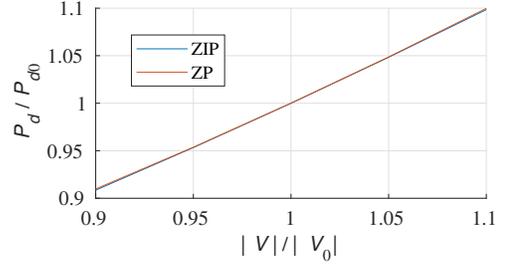} 
  \vspace{-3mm}
  \caption{\textcolor[rgb]{0,0,0}{Comparison between ZIP and ZP model based on the data collected from a real-life distribution system.} \vspace{-3mm}}\label{ZIPZPfig}
\end{figure}

\textcolor[rgb]{0,0,0}{Moreover, only a small share of distribution loads is constant current.
Here the accuracies of ZP and ZIP models are compared in practice based on the minute-by-minute data collected from the load points of a real-life distribution system in Dublin, Ireland \cite{McKenna2017}.
Coefficients $a$, $b$, and $c$ are estimated for every minute of a sample day at all load points, using the high-resolution measured data records.
Fig. \ref{ZIPZPfig} presents the average of $P_d^t$/${P_d}_0^t$ found based on both ZIP and ZP models for $t$=1:1440 minutes. Considering the same chance for all possible values of $\left|V\right|$/$V_0$ in the range of [0.9 1.1], the standard deviation of normalized active powers calculated using ZP model compared with the ZIP counterparts is 0.057\%. The maximum deviation of active powers calculated with ZP model with respect to the ZIP counterparts is 0.17\%. With the minute-by-minute measured voltages taken into account, these standard and maximum deviations are 0.021 and 0.028\%, respectively.
However, if the level of accuracy that ZP model or even ZIP model provides is not sufficient, more accurate voltage-dependent load models can be applied. The proposed method can easily accommodate any load model with desired level of complexity and accuracy.}

The voltage-dependent behavior is not fixed and the model should be kept up to date.
\textcolor[rgb]{0,0,0}{Coefficients $a'_p$ and $c'_p$ and $a'_q$ and $c'_q$ are not independent, i.e., $a'_p+c'_p=a'_q+c'_q=1$. Therefore, in order} to update this model for $t$=$\tau$+1, four coefficients should be determined,
i.e., $a'_p$, $P_{d_0}$, $a'_q$, and $Q_{d_0}$. Two sets of measurements for $P_d$, $Q_d$ and $V$ are required. The first/second set of parameters is measured before/after applying the changes proposed by the scheduling framework.
The following algorithm updates the model for $t$=$\tau$+1.

\begin{enumerate}
	\item Consider the values forecasted for or the last update of $a'_p$ and $a'_q$.
	\item \textcolor[rgb]{0,0,0}{Measure/estimate $\left|V\right|$, $P_d$ and $Q_d$ and calculate $P_{d_0}$ and $Q_{d_0}$ using \eqref{ZP}.}
	\item Solve the \textsc{dsp} with this load model for $t$=$\tau$+1, and the load model given in the scenarios for $t=\tau+2$:$\tau+T$.
	\item Apply the set-points proposed.
	\item Measure/estimate $\left|V\right|$, $P_d$ and $Q_d$.
	\item \textcolor[rgb]{0,0,0}{Two sets of measurements are now available from steps 2 and 5. Use the first equation of \eqref{ZP} with these two set of measurements to update $a'_p$ and $P_{d_0}$. Use the second equation of \eqref{ZP} to update $a'_q$ and $Q_{d_0}$.}
	\item If the change in the updated values is not negligible, solve the \textsc{dsp} again using the updated load model coefficients.
\end{enumerate}

\textcolor[rgb]{0,0,0}{The values of $\left|V\right|$, $P_d$ and $Q_d$ are not measured at all buses. In case that there are enough measurements, state estimation techniques, such as those reviewed in \cite{Dehghanpour2019}, can be applied to approximate these values for the buses at which these parameters are not measured.
If this is not the case, the load model coefficients for period $\tau+1$ should also be included in the scenario vectors in the same way as periods $\tau$+1:$\tau$+$T$.
The method that is presented here for updating the load model for period $\tau+1$ can still be used to reduce the degree of uncertainties on these parameters.
Using the limited number of measurements before and after applying the optimal changes, historical data, system physical equations and equations \eqref{ZP}, state estimation techniques can approximate the mean values and standard deviations of the load model coefficients in period $\tau+1$. These statistical moments can be used to extract the scenarios on these coefficients along with the other uncertain parameters using the method presented in subsection \ref{Uncertain}.}

\textcolor[rgb]{0,0,0}{This paper deals with the \textsc{dsp} and steady state voltage control problem.
In the real time operation of distribution systems, some other effects, e.g., dynamic voltage stability and harmonic distortion issues, may need to be taken into account.
To deal with such issues, dynamic and nonlinear load models should be applied, respectively.
Dynamic and non-linear control techniques and load models cannot be used for steady state voltage control.
In the \textsc{dsp} and steady state voltage control problem, the voltage-dependent behavior of loads needs to be taken into account.
ZIP model is applied to formulate the nonlinear relationship between the load and voltage levels with an acceptable accuracy.
The loads can be of any types. The ZIP coefficients at each load point depend on many factors and change from time to time. For instance, the ZIP coefficients of a motor load depend on motor type, size, load and speed.
For every load point, the coefficients are estimated for period $\tau+1$ and are included in the scenarios for $\tau+2$:$\tau+T$.}

\textcolor[rgb]{0,0,0}{The equation of apparent power demand at each network bus is given in \eqref{APRPOW}. The negative sign indicates that $P_d$ and $Q_d$ are consumed.
Combining the perturbed counterparts of \eqref{ZP} and \eqref{APRPOW}, the perturbed load model is given in \eqref{Load_Linear_2}.}
\begin{equation}
\label{ZIP}
\begin{array}{l}
\hspace{-3.5mm} 
\displaystyle
\frac{P_d}{P_{d_0}} \hspace{-1.25mm} = \hspace{-1.00mm} a_p\frac{|V|^2}{V_0^2} \hspace{-1mm} + \hspace{-0.75mm} b_p\frac{|V|}{V_0} \hspace{-1mm} + \hspace{-0.75mm} c_p,

\hspace{3mm}

\frac{Q_d}{Q_{d_0}} \hspace{-1.25mm} = \hspace{-1.00mm} a_q\frac{|V|^2}{V_0^2} \hspace{-1mm} + \hspace{-0.75mm} b_q\frac{|V|}{V_0} \hspace{-1mm} + \hspace{-0.75mm} c_q
\end{array}
\end{equation}
\begin{equation}
\label{ZP}
\begin{array}{l}

\displaystyle
\frac{P_d}{P_{d_0}}=a'_p \frac{|V|^2}{V_0^2}+c'_p, 

\hspace{7mm}

\frac{Q_d}{Q_{d_0}}=a'_q\frac{|V|^2}{V_0^2}+c'_q

\end{array}
\end{equation}
\begin{equation}
\label{APRPOW}
\begin{array}{l}

\displaystyle
\textcolor[rgb]{0,0,0}{P_d+\text{j}Q_d=-VI^*}

\end{array}
\end{equation}
\begin{equation}
\label{Load_Linear_2}
\begin{array}{l}
\displaystyle
\hspace{-4.75mm}
\begin{pmatrix} \hspace{-0.5mm} \Delta{I}_x \hspace{-0.5mm} \\ \hspace{-0.5mm} \Delta{I}_y \hspace{-0.5mm} \end{pmatrix} \hspace{-1.5mm}
= \hspace{-1.5mm}
\begin{pmatrix} \hspace{-0.5mm} \hat{V_x} \hspace{-2mm} & \hspace{-2mm} \hat{V_y} \hspace{-1mm} \\ \hspace{-0.5mm} \hat{V_y} \hspace{-2mm} & \hspace{-2.5mm} -\hat{V_x} \hspace{-1mm} \end{pmatrix}^{\hspace{-1.75mm} -1} \hspace{-2mm}
\vspace{1mm}

\begin{pmatrix}
\hspace{-0.5mm}
\frac{2a'_p P_{d_0} \hspace{-0.5mm} \hat{V_x}}{V_0^2} \hspace{-1mm} - \hspace{-1mm} \hat{I_x} \hspace{-1.5mm}
& \hspace{-1.5mm} \frac{2a'_p P_{d_0} \hspace{-0.5mm} \hat{V_y}}{V_0^2} \hspace{-1mm} - \hspace{-1mm} \hat{I_y} \hspace{-1.5mm}  \\

  \hspace{-1mm} \frac{2a'_q Q_{d_0} \hspace{-0.5mm} \hat{V_x}}{V_0^2} \hspace{-1mm} + \hspace{-1mm} \hat{I_y} \hspace{-2mm}
& \hspace{-1.5mm} \frac{2a'_q Q_{d_0} \hspace{-0.5mm} \hat{V_y}}{V_0^2} \hspace{-1mm} - \hspace{-1mm} \hat{I_x} \hspace{-1.5mm}
\end{pmatrix} \hspace{-2mm}

\begin{pmatrix} \hspace{-1mm} \Delta{V}_x  \hspace{-1mm} \\ \hspace{-1mm} \Delta{V}_y \hspace{-1mm} \end{pmatrix}
\end{array}
\hspace{-3.95mm}
\end{equation}
\vspace{-5mm}
%
\subsection{Objective Function and Perturbed Constraints}\label{SecIIIB}
Here, the objective is to minimize the system cost for the deterministic single-period \textsc{dsp} \textcolor[rgb]{0,0,0}{and is given in \eqref{OBJ_Deter}}.
\textcolor[rgb]{0,0,0}{It is shown in Section \ref{Stoch_Alg} how to extend this formulation to achieve the formulation of stochastic \textsc{dsp}.
The objective function includes the change in the operation cost ($\Delta OC_t$) and control cost ($CC_t$) due to changing the control variables from $\hat{\upsilon}_t$ to $\upsilon_t$.}
$\Delta OC_t$ and $CC_t$ are presented in \eqref{OBJ_OC} and \eqref{OBJ_CC}, respectively. The voltage constraints are modeled as soft constraints, i.e., the voltage deviations are penalized in the objective function.

Binary variables $u_{k_{TC}}$ and $u_{k_{CB}}$ should satisfy \eqref{tap_change} and \eqref{cap_change}.
\textcolor[rgb]{0,0,0}{In \eqref{tap_change}, $\Delta tap^{max}_{k_{TC}}$ is the maximum possible tap change. If such parameter does not exist for the transformer, $\Delta tap^{max}_{k_{TC}}=\overline{tap_{k_{TC}}}-\underline{tap_{k_{TC}}}$.}
The perturbed voltage constraints at bus $b$ are given in \eqref{VolCon0}.
The current constraint of line $l$ is given in \eqref{CurCon0}. Neglecting the small term (${\Delta V_1}_x$-${\Delta V_2}_x$)(${\Delta V_1}_y$-${\Delta V_2}_y$), a perturbed formulation for the maximum current constraint is provided in \eqref{CurCon}.
For simplicity, the index $t$ has been removed from \eqref{VolCon0} and \eqref{CurCon}.
The sending and receiving buses are given by subscripts 1 and 2, respectively.
\textcolor[rgb]{0,0,0}{As shown in \eqref{TAPCAPLim}}, the tap positions and \textsc{cb} steps should be set within their limits.

\textcolor[rgb]{0,0,0}{It is imperative to distinguish between $\hat{\upsilon}_t$ and $\upsilon_{t-1}$. For instance, in \eqref{tap_change}, $\hat{tap}_{k_{TC},t}$ is the tap position of transformer $k_{TC}$ in period $t$ that is found in the previous sub-problem of \textsc{tra} or is included in the initial guess. $tap_{k_{TC},t-1}$ is the tap position in the previous period.
The model function and perturbed constraints are built around $\hat{\upsilon}_t$.}
\begin{equation}
\label{OBJ_Deter}
\begin{array}{l}
\displaystyle
\hspace{-3.5mm}
\underbrace{Min}_{\upsilon_t} \hspace{1mm}
\left\{{F_D}_t\right\} \hspace{-1mm}
=\hspace{-1mm}
\sigma_t \Delta OC_t \hspace{-1mm}
+ \hspace{-0.5mm}
CC_t+ \hspace{-1mm} \sum_{b=1}^{N_b} \eta_b \left[\underline{{\vartheta_b}_t}+\overline{{\vartheta_b}_t}\right]
\end{array}
\hspace{-2mm}
\end{equation}
\vspace{-2mm}
\begin{equation}
\label{OBJ_OC}
\begin{array}{l}
\displaystyle
\Delta OC_t=
{\rho_A}_t {\Delta P_p}_t
+
{\rho_R}_t {\Delta Q_p}_t
+  \hspace{-3mm}
\sum_{e\in{DER}}{\rho_e{\Delta P_g}_{e,t}}
\end{array}
\end{equation}
\vspace{-1mm}
\begin{equation}
\label{OBJ_CC}
\begin{array}{l}
\displaystyle
CC_t=
\sum_{k_{TC} \in OLTC} \hspace{-2mm} u_{k_{TC},t} \textsl{D}_{k_{TC}}
+ \hspace{-3mm}
\sum_{k_{CB} \in CB} \hspace{-2mm} u_{k_{CB},t} \textsl{D}_{k_{CB}}
\end{array}
\end{equation}
\begin{equation}
\label{tap_change}
\begin{array}{l}

\displaystyle
\hspace{-2mm}
\Delta tap_{k_{TC},t} \hspace{-0.75mm} + \hspace{-0.75mm} \hat{tap}_{k_{TC},t} \hspace{-0.75mm} - \hspace{-0.75mm}  tap_{k_{TC},t-1} \hspace{-1mm} \leq \hspace{-1mm} u_{k_{TC},t}  \Delta tap^{max}_{k_{TC}}
\vspace{1mm}
\\
\displaystyle
\hspace{-2mm}
-\Delta tap_{k_{TC},t} \hspace{-0.75mm} - \hspace{-0.75mm} \hat{tap}_{k_{TC},t} \hspace{-0.75mm} + \hspace{-0.75mm} tap_{k_{TC},t-1} \hspace{-1mm} \leq \hspace{-1mm}  u_{k_{TC},t}  \Delta tap^{max}_{k_{TC}}

\end{array}
\end{equation}
\begin{equation}
\label{cap_change}
\begin{array}{l}

\displaystyle
\Delta st_{k_{CB},t} \hspace{-0.5mm} + \hspace{-0.5mm} \hat{st}_{k_{CB},t} \hspace{-0.5mm} - \hspace{-0.5mm}  st_{k_{CB},t-1} \leq  u_{k_{CB},t}  \Delta st^{max}_{k_{CB}}
\vspace{1mm}
\\
\displaystyle
-\Delta st_{k_{CB},t} \hspace{-0.5mm} - \hspace{-0.5mm} \hat{st}_{k_{CB},t} \hspace{-0.5mm} + \hspace{-0.5mm} st_{k_{CB},t-1} \leq  u_{k_{CB},t}  \Delta st^{max}_{k_{CB}}

\end{array}
\end{equation}
\begin{equation}
\label{VolCon0}
\begin{array}{l}
\displaystyle
\hspace{-4mm}
\underline{V_b}^2 \hspace{-1.5mm}
- \hspace{-1mm}
\left|\hat{V_b}\right|^2 \hspace{-1mm}
- \hspace{-1mm}
\underline{\vartheta_b}
 \hspace{-0.5mm}
\leq \hspace{-0.5mm}
2{\hat{V_b}}_x\Delta{V_b}_x \hspace{-1.5mm}
+ \hspace{-1mm}
2{\hat{V_b}}_y\Delta{V_b}_y \hspace{-0.5mm}
\leq \hspace{-0.5mm}
\overline{V_b}^2 \hspace{-1.5mm}
- \hspace{-1mm}
\left|\hat{V_b}\right|^2 \hspace{-0.5mm}
+ \hspace{-0.5mm}
\overline{\vartheta_b}
\end{array}
\hspace{-3mm}
\end{equation}
\begin{equation}
\label{CurCon0}
\begin{array}{l}
\displaystyle

 \left|V_1-V_2\right|^2 \leq \hspace{-1mm}
(R_l^2+X_l^2)
\overline{I_l}^2

\end{array}
\hspace{-3mm}
\end{equation}
\begin{equation}
\label{CurCon}
\begin{array}{l}
\displaystyle

\hspace{-2mm} (\hat{V_1}_x \hspace{-1mm} - \hspace{-1mm} {\hat{V_2}}_x)\Delta {V_1}_x      \hspace{-1mm}   +  \hspace{-1mm}  ({\hat{V_1}}_x \hspace{-1mm} - \hspace{-1mm} {\hat{V_2}}_x)\Delta {V_2}_x\\
\displaystyle

\hspace{-4.5mm} + (\hspace{-0.25mm} {\hat{V_1}}_y \hspace{-1.25mm} - \hspace{-1.0mm} {\hat{V_2}}_y)\Delta \hspace{-0.5mm} {V_1}_y      \hspace{-1.25mm}   +  \hspace{-1mm}  ( \hspace{-0.5mm} {\hat{V_1}}_y \hspace{-1mm} - \hspace{-1mm} {\hat{V_2}}_y)\Delta \hspace{-0.25mm} {V_2}_y

\hspace{-1mm} \leq \hspace{-1mm}
(R_l^2 \hspace{-1mm} + \hspace{-1mm} X_l^2)
\frac{\overline{I_l}^2 \hspace{-1mm}
- \hspace{-1mm}
\left|\hat{I_l}\right|^2}{2}

\end{array}
\hspace{-3mm}
\end{equation}
\begin{equation}
\label{TAPCAPLim}
\begin{array}{l}
\displaystyle

\textcolor[rgb]{0,0,0}{\overline{tap_{k_{TC}}} \leq \hat{tap}_{k_{TC},t}+\Delta tap_{k_{TC},t} \leq \overline{tap_{k_{TC}}}} \\
\displaystyle
\textcolor[rgb]{0,0,0}{\hspace{3mm} \underline{st_{k_{CB}}} \leq \hat{st}_{k_{CB},t}+\Delta st_{k_{CB},t} \leq \overline{st_{k_{CB}}}}

\end{array}
\end{equation}

The perturbed voltages and currents in this formulation are rewritten in terms of the optimization variables by combining the models extracted for the controllable devices and system loads with the network model \textcolor[rgb]{0,0,0}{provided in \eqref{NetEq}}.
The non-perturbed notations are constant.
\textcolor[rgb]{0,0,0}{The perturbed currents in \eqref{NetEq} are replaced by the perturbed control variables and voltages using the perturbed equations presented for the controllable devices and loads. The resultant equations would be \eqref{LinModel0} and \eqref{LinModel}.} $\Upsilon$ and $N_\upsilon$ are the vector and number of control variables, respectively. Matrices $\boldsymbol{A}$ and $\boldsymbol{B}$ are found using matrices $A$ and $B$ developed for the controllable devices and loads.

As discussed in subsection \ref{Sol}, to apply \textsc{tra}, all inequality constraints
should be converted to equality constraints using auxiliary variables. Simple bounds on the optimization variables can also be handled. For instance, \eqref{VolCon0} is converted to \eqref{VolCon0C} using positive auxiliary variables $\overline{\epsilon_b}$ and $\underline{\epsilon_b}$.
\begin{equation}
\label{NetEq}
\begin{array}{l}
\displaystyle
\hspace{-1mm}
\begin{pmatrix} \Delta{I}_x \\ \Delta{I}_y \end{pmatrix}_{2N_b\times{1}} \hspace{-1.5mm}
=\hspace{-1mm}
\underbrace{\begin{pmatrix} {Y_\text{Bus}}_x & -{Y_\text{Bus}}_y
\\
{Y_\text{Bus}}_y & {Y_\text{Bus}}_x \end{pmatrix}}_{\boldsymbol{Y}} \hspace{-0.5mm}
.\hspace{-1mm}
\begin{pmatrix} \Delta{V}_x \\ \Delta{V}_y \end{pmatrix}_{2N_b\times{1}} \hspace{-2mm}
\end{array}
\end{equation}
\vspace{-2mm}
\begin{equation}
\label{LinModel0}
\begin{array}{l}
\displaystyle
\hspace{-3mm}
\boldsymbol{A}
 \hspace{-1mm}
\begin{pmatrix} \Delta{V}_x \\ \Delta{V}_y \end{pmatrix}
+\hspace{-1mm}
\left[\boldsymbol{B}\right]_{2N_b\times{2N_\upsilon}}\left[{\Upsilon}\right]_{2N_\upsilon \times{1}} 
=\hspace{-1mm}
\boldsymbol{Y}
\hspace{-1mm}
\begin{pmatrix} \Delta{V}_x \\ \Delta{V}_y \end{pmatrix} \hspace{-2mm}
\end{array}
\end{equation}
\vspace{-2mm}
\begin{equation}
\label{LinModel}
\begin{array}{l}
\displaystyle
\hspace{-5mm}
\begin{pmatrix} \Delta{V}_x \\ \Delta{V}_y \end{pmatrix}
=
\left(\boldsymbol{Y}-\boldsymbol{A}\right)^{-1}\boldsymbol{B}.\Upsilon 
\end{array}
\end{equation}
\begin{equation}
\label{VolCon0C}
\begin{array}{l}
\displaystyle
\hspace{-2mm}
2{V_b}_x\Delta{V_b}_x \hspace{-0.75mm} + \hspace{-0.75mm} 2{V_b}_y\Delta{V_b}_y \hspace{-0.75mm} - \hspace{-0.75mm} \overline{\vartheta_b} \hspace{-0.75mm}
+ \hspace{-0.75mm} \overline{\epsilon_b} \hspace{-0.75mm} = \hspace{-0.75mm} \overline{V_b}^2 \hspace{-0.75mm} - \hspace{-0.75mm} \left|V_b\right|^2, \hspace{3mm} \overline{\epsilon_b} \geq 0
\vspace{1mm}
\\
\displaystyle
\hspace{-2mm}
2{V_b}_x\Delta{V_b}_x \hspace{-0.75mm} + \hspace{-0.75mm} 2{V_b}_y\Delta{V_b}_y \hspace{-0.75mm} + \hspace{-0.75mm} \underline{\vartheta_b} \hspace{-0.75mm}
- \hspace{-0.75mm} \underline{\epsilon_b} \hspace{-0.75mm} = \hspace{-0.75mm} \underline{V_b}^2 \hspace{-0.75mm} -\hspace{-0.75mm} \left|V_b\right|^2, \hspace{3mm} \underline{\epsilon_b} \geq 0
\end{array}
\end{equation}

In \textsc{tra} sub-problems, the model function is a quadratic function of perturbed optimization variables.
In \eqref{OBJ_OC}, ${\Delta P_p}_t$ and ${\Delta Q_p}_t$ are non-linear in terms of optimization variables.
\textcolor[rgb]{0,0,0}{${\Delta P_p}_t$ and ${\Delta Q_p}_t$ can be replaced by ${\Delta V_p}_t$ and ${\Delta I_p}_t$ using ${P_p}_t$+j${Q_p}_t$=${V_p}_t{I_p}_t^*$. ${V_p}_t$ is the voltage at and ${I_p}_t$ is the current injected to the primary side of transformer. Using the model developed for \textsc{oltc} transformers, ${\Delta V_p}_t$ and ${\Delta I_p}_t$ are rewritten in terms of ${\Delta V_s}_t$ and ${\Delta I_s}_t$ and $\Delta tap_t$. ${\Delta V_s}_t$ and ${\Delta I_s}_t$ are rewritten in terms of the perturbed control variables using \eqref{LinModel}.}
\section{Stochastic Scheduling}\label{Stoch_Alg} 
The objective function of the proposed stochastic scheduling algorithm is given in \eqref{OBJ_Stoch}.
To make a compromise between the solution accuracy and computational burden, $\sigma_t$ is increased as $t$ increases.
\textcolor[rgb]{0,0,0}{In a day-ahead formulation for the \textsc{dsp}, in every time period, the \textsc{oltc}(s') tap position(s) are the same for all scenarios.
However, in the proposed \textsc{rhc}-based method,
only the tap positions found for the first period ($\tau$+1) are the same for all scenario.
The reason is that only the decisions made for this period are really applied.}
Therefore, except for $tap_{\tau+1}$ the tap positions for other periods take subscript $s$.

More details were provided in Section \ref{Prop}.
All the constraints introduced for the deterministic single period \textsc{dsp} should also be included.
Constraint \eqref{tap_change} is rewritten as \eqref{tap_change2}-\eqref{tap_change3},
where the point $\hat{\upsilon}$ is the initial point to build the perturbed model (or the solution of the previous \textsc{tra} sub-problem) around it.
Subscript $k_{TC}$ is removed for brevity.
Constraint \eqref{cap_change} is rewritten similarly.
Only the decisions taken for period $\tau$+1 are implemented
and the proposed algorithm is applied again to extract the optimal schedule of the next period.
The number of tap changing operations is limited using \eqref{tap_change5}. For \textsc{cb}s, the same constraints are considered.
\vspace{-1mm}
\begin{equation}
\label{OBJ_Stoch}
\begin{array}{l}
\displaystyle
\hspace{-2.5mm}

\underbrace{Min}_{\Upsilon}
\hspace{+1.5mm}
\left\{{F_S}_\tau \right\} 
=
\sum_{s=1}^{N_s} \pi_s

\sum_{t=\tau+1}^{\tau+T} \hspace{-1mm} \left( \sigma_t  \Delta OC_{t,s} \hspace{-0.25mm} + \hspace{-0.25mm} CC_{t,s} \right)
\\
\displaystyle
\hspace{30mm} + \sum_{s=1}^{N_s} \pi_s \sum_{b=1}^{N_b} \hspace{-0.25mm} \eta_b \left[\underline{{\vartheta_b}_{t,s}}+\overline{{\vartheta_b}_{t,s}}\right]\\
\displaystyle
\textcolor[rgb]{0,0,0}{\Upsilon=\left\{\upsilon_{\tau+1},\hspace{2mm} \upsilon_{\tau+1,s},\hspace{2mm} \upsilon_{t,s} \hspace{3mm} {\forall \ \tau+2 \leq t \leq T+\tau}\right\}}\\

\end{array}
\hspace{-2mm}
\vspace{-1mm}
\end{equation}
\begin{equation}
\label{tap_change2}
\begin{array}{l}
\displaystyle
\forall s,t \geq \tau+2 \\

\displaystyle
\hspace{-3.5mm} \Delta tap_{s,t} \hspace{-0.5mm}  - \hspace{-0.5mm} \Delta tap_{s,t-1} \hspace{-0.5mm} +  \hspace{-0.5mm}
\hat{tap}_{s,t} \hspace{-0.5mm} - \hspace{-0.5mm}  \hat{tap}_{s,t-1} \hspace{-1mm} \leq  \hspace{-1mm} u_{s,t}  \Delta tap^{max} \hspace{-2mm} \\

\displaystyle
\hspace{-3.5mm} \Delta tap_{s,t-1} \hspace{-0.5mm} - \hspace{-0.5mm} \Delta tap_{s,t} \hspace{-0.5mm} -  \hspace{-0.5mm}
\hat{tap}_{s,t} \hspace{-0.5mm} + \hspace{-0.75mm} \hat{tap}_{s,t-1} \hspace{-1mm} \leq \hspace{-1mm}  u_{s,t}  \Delta tap^{max} \hspace{-2mm}

\end{array}
\end{equation}
\begin{equation}
\label{tap_change3}
\begin{array}{l}
\displaystyle
\Delta tap_{\tau+1} + \hat{tap}_{\tau+1} -  tap_{\tau} \leq  u_{\tau+1}  \Delta tap^{max} \\
\displaystyle
-\Delta tap_{\tau+1} - \hat{tap}_{\tau+1} +  tap_{\tau} \leq  u_{\tau+1}  \Delta tap^{max}
\end{array}
\end{equation}
\begin{equation}
\label{tap_change5}
\begin{array}{l}
\displaystyle
u_{k_{TC},\tau+1} + \sum_{t=\tau+2}^{T+\tau} u_{k_{TC},s,t} \leq \overline{N_{TC_T}} \hspace{7mm}  \forall s
\end{array}
\end{equation}
\section{Case Studies}\label{Case_Studies}
\begin{figure}[!t]
  \centering
  \includegraphics*[width=0.98\columnwidth,trim=0.3cm 2.8cm 1.38cm 0.89cm,clip=true]{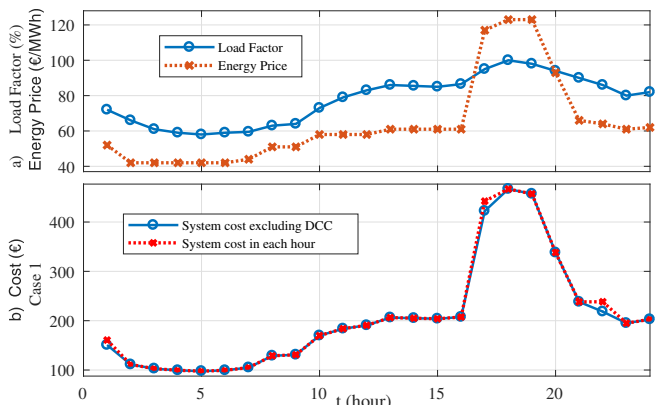}
	\vspace{-5mm}
  \caption{a) Load factors and energy prices, b) system costs in Case 1.}\label{LoadCost}
	\vspace{-3mm}
\end{figure}

\begin{figure}[!t]
  \centering
  \includegraphics*[width=0.97\columnwidth,trim=0.5cm 2.90cm 0.5cm 0.5cm,clip=true]{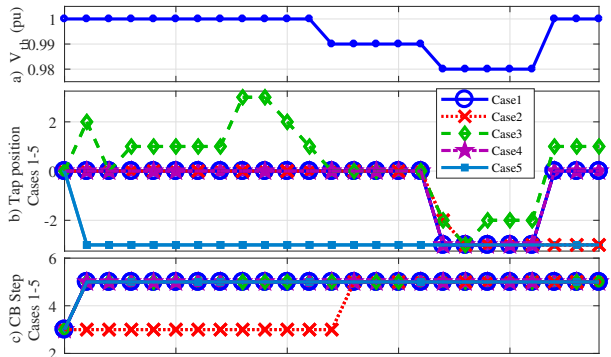}
	\vspace{-3.5mm}
  \caption{a) Upstream system voltages, b) \textsc{oltc} tap positions, c) \textsc{cb} steps.}\label{TapSt}
\end{figure}

The proposed algorithm is tested on IEEE 33-bus test system.
Table \ref{CDs} presents the data of two parallel \textsc{oltc} transformers which connect this system to the upstream network and
a \textsc{der}, a \textsc{pv} unit, a \textsc{cb} and an \textsc{svr} which are added to this system.
The average hourly load levels and energy prices
are presented in Fig. \ref{LoadCost}(a) and $\rho_R$=0.2$\rho_A$.
The average coefficients of ZIP model is found in \cite{Nouri2017}.
\textcolor[rgb]{0,0,0}{The PV unit produces active power from 7 AM to 7 PM. The maximum producible power ($\overline{S}_{PV}$) happens at 1 PM. From 6 AM to 1 PM and from 1 PM to 8 PM, ${P_g}^{\textsc{PV}}$ changes linearly from zero to $\overline{S}_{PV}$ and from $\overline{S}_{PV}$ to 0, respectively.}
Fig. \ref{TapSt} (a) shows the average upstream Thevenin voltage levels. \textcolor[rgb]{0,0,0}{$\overline{V}$=1.05, $\underline{V}$=0.95 and $\sigma$=1 h.}
\newcolumntype{N}[1]{>{\centering\let\newline\\\arraybackslash\hspace{0pt}}m{#1}}
\newcolumntype{C}[1]{>{\centering\let\newline\\\arraybackslash\hspace{0pt}}m{#1}}
\newcolumntype{X}[1]{>{\centering\let\newline\\\arraybackslash\hspace{-2pt}}m{#1}}
\newcolumntype{R}[1]{>{\centering\let\newline\\\arraybackslash\hspace{-2pt}}m{#1}}
\newcolumntype{M}[1]{>{\centering\let\newline\\\arraybackslash\hspace{0pt}}m{#1}}
\newcolumntype{I}[1]{>{\centering\let\newline\\\arraybackslash\hspace{0pt}}m{#1}}
\newcolumntype{H}[1]{>{\centering\let\newline\\\arraybackslash\hspace{-2pt}}m{#1}}
\newcolumntype{L}[1]{>{\centering\let\newline\\\arraybackslash\hspace{-2pt}}m{#1}}
\newcolumntype{U}[1]{>{\centering\let\newline\\\arraybackslash\hspace{-2pt}}m{#1}}

So far, the average values ($\mu$) of all uncertain parameters have been presented.
A forecast error of $e_{t,\xi}$ is assigned to uncertain parameter $\xi$ at period $t$.
For the first period ($t=\tau+1$),
$e_{\tau+1,\xi}=0.02\mu_{\tau+1,\xi}$. This forecast error increases linearly for the upcoming periods with the rate of $0.1e_{\tau+1,\xi}$ per period.
The uncertainty of the parameters is modeled using a joint normal \textsc{pdf} with the average values of $\mu_{t}$ and standard deviations of $e_{t}$.
The correlation between the load levels at each two buses at a certain period is considered to be 80\%. The correlation between the load levels at a certain bus and $V_{th}$, $Z_{th}$, active power prices and reactive power prices at successive time periods are considered to be 70, 90, 80, 90 and 90\%, respectively. Other correlations are not taken into account.

The proposed algorithm is applied from 1 AM
with an infeasible starting point
($tap_1$=$tap_2$=0, $st$=3, $Pg^{\text{DER}}$=$Qg^{\text{DER}}$=$Qg^{\text{PV}}$=0 and $Qg^{\text{SVR}}$=300 kVAR).
In all studies, the proposed framework is applied 24 times. In the first study the \textsc{sh} ($T$) is 20 hours.
It can be assumed that a 20-hour scheduling window is being moved hour by hour to solve the \textsc{dsp},
considering the system future. For some hours, a certain length of this window happens in the next day with all parameters same as those specified for the current day.
\begin{table}[!t]
    \caption{Controllable Devices and \textsc{dsp} parameters}
		\vspace{-2mm}
		\label{CDs}
		\centering
		\setlength\tabcolsep{1.5pt} 
		\setlength{\doublerulesep}{2\arrayrulewidth}
\begin{tabular}{c| c | c }
\hline
 & Bus & Characteristics \\
\hline\hline
DER & 14 & $\overline{S}$=500 kVA, $\overline{P_g}$=500 kW, $\overline{Q_g}$=50 kVAR, $\rho$=60 \euro/MWh \\
\hline
PV & 14 & $\overline{S}=250$ kVA, $\overline{P_g}=250$ kW, $\overline{\alpha}^{pv}=35^\text{o}$ \\
\hline
SVR & 30 & $\overline{Q_g}=500$ kVAR \\
\hline
CB & 33 & $\overline{Q_g}=500$ kVAR, $\overline{st}=5$ \\
\hline
\multicolumn{3}{c}{$Z^n_{sr}$=$Z_{th}$=0.01+0.05j, $\overline{tap}$=3, $\underline{tap}$=-3, $\Delta{U}$=1\%, $R_c^n$=100 and $X_M^n$=95 pu.} \\
\multicolumn{3}{c}{$T$=20 h, $\sigma$=1 h, $D_{OLTC}$=\euro20, $D_{CB}$=\euro10, $\overline{V}$=1.05, $\underline{V}$=0.95} \\
\hline
\end{tabular}
\vspace{-3mm}
\end{table}
\begin{table}[!t]
    \caption{Case Studies}
		\vspace{-2mm}
		\label{Cases}
		\centering
		\setlength\tabcolsep{4.5pt} 
		\setlength{\doublerulesep}{2\arrayrulewidth}
\begin{tabular}{ N{0.5cm} | C{0.6cm} | X{1cm} | R{0.6cm} | M{0.3cm} | I{0.8cm} | H{0.58cm}| L{0.75cm} | U{0.7cm}}
\hline
Case & Load Model & $D_{OLTC}$ (\euro) & $D_{CB}$ (\euro) & $T$ (h) & OC (\euro) & DCC (\euro) & Cost (\euro) & Loss (MWh)  \\
\hline
\hline
1 & ZIP & 20  & 10 & 20 & 4931.4 & 50.0 & 4981.4 & 2.11\\
\hline
2 & ZIP & 20  & 10 & 1 & 4973.3 & 50.0 & 5023.3 & 2.36\\
\hline
3 & ZIP & -  & - & 20 & 4883.9 & 190.0 & 5073.9 & 2.12\\
\hline
4 & ZIP & 40  & 20 & 20 & 4931.4 & 100 & 5031.4 & 2.11 \\
\hline
5 & P & 20  & 10 & 20 & 5136.3 & 30.0 & 5166.3 &  2.96 \\
\hline
0{$^*$} & ZIP & 20  & 10 & 20 & 5032.1 & 50.0 & 5082.1 &  2.22 \\
\hline
\end{tabular}
\vspace{+0.25mm}
\\
\footnotesize{$^*$ Case 0 with Inaccurate Upstream Model}\\
\end{table}
\begin{figure}[t!]
  \centering
\includegraphics[clip = true,width=0.91\columnwidth,trim=0.5cm 0.12cm 0.5cm 0.3cm,]{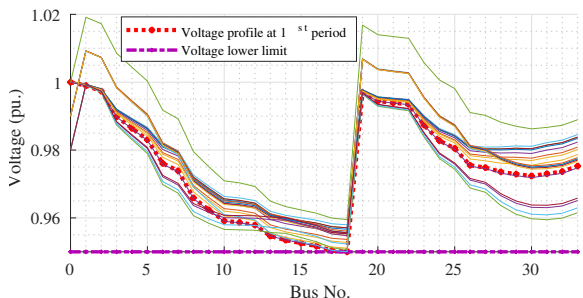}
\vspace{-3mm}
  \caption{\textcolor[rgb]{0,0,0}{Voltage profiles for $t$=1-20 in the first scheduling window.}}
\label{Voltage1}
\vspace{-2.5mm}
\end{figure}

\begin{figure}[!t]
  \centering
  \includegraphics*[width=0.99\columnwidth,trim=0.43cm 0.05cm 0.50cm 0.3cm,clip=true]{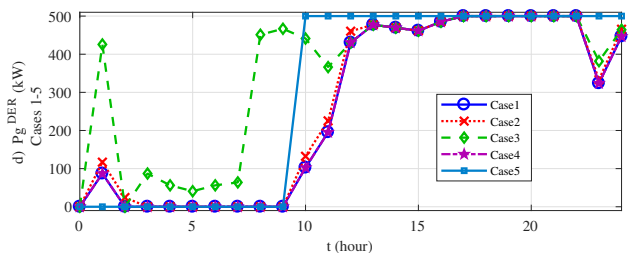} \vspace{-3mm}
	\vspace{-1mm}
  \caption{Power productions of the dispatchable \textsc{der}.}\label{DERPg}
	\vspace{-3.75mm}
\end{figure}

\begin{figure}[!t]
  \centering
  \includegraphics*[width=0.95\columnwidth,trim=0.5cm 0.1cm 0.02cm 0.1cm,clip=true]{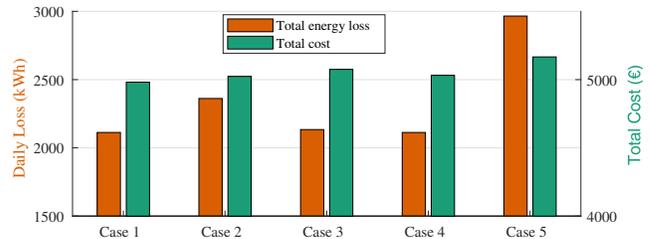}
	\vspace{-3mm}
  \caption{Energy loss and total cost in Cases 1-5.}\label{CostLoss}
	\vspace{-1.25mm}
\end{figure}

\textcolor[rgb]{0,0,0}{Six different studies, i.e., cases 1-5 and case 0, are conducted (see Table \ref{Cases}).
In cases 1-5, the results of applying the proposed algorithm are provided in Figs. \ref{LoadCost}-\ref{CostLoss}.
Fig. \ref{LoadCost} presents the total costs and discrete control costs (\textsc{dcc}s) in Case 1 along with the average energy prices and load factors.
Fig. \ref{Voltage1} gives the voltage profiles in case 1 for hours $t$=1:20, i.e, the first scheduling window, for the most probable scenario.
Each voltage profile corresponds to one of the periods $t$=1:20. The main purpose of this figure is to show the voltages never violate the limits and to identify the bus with the minimum voltage level.
Fig. \ref{TapSt} shows the \textsc{dca}s to analyze the effects of changing the parameters of the proposed algorithm.
In this figure, the variations of the average Thevenin voltages of the upstream system are also presented.
The production levels of the \textsc{der} are presented in Fig. \ref{DERPg}.
The results provided in Figs. \ref{DERPg} and \ref{TapSt} show the interconnections between the discrete and continuous control actions in the simultaneous scheduling of fast and slow control devices.
Fig. \ref{CostLoss} compares the total costs and energy losses.
These total costs and energy losses are also presented in Table \ref{Cases} along with the operation costs (\textsc{oc}s) and \textsc{dcc}s.}

\textbf{In Case 1}, at the first period,
the framework proposes the cheapest actions. The \textsc{cb} step is increased by two steps in Fig. \ref{TapSt} (c) and $P_g^{der}$ is raised by 80 kW in Fig. \ref{DERPg}.
\textcolor[rgb]{0,0,0}{These control actions solve the under-voltage issues for the initial time periods. Meanwhile, the voltage profiles are kept as low as possible, of course, without violating the lower bound voltage limit. This reduces the power demand considering the voltage-dependent nature of the loads.
Fig. \ref{Voltage1} presents the voltage profiles for all periods ($t$=1:20) in the first scheduling window for the most probable scenario.
The voltage at bus 18 hits the minimum level in some initial periods ($V_{t=1,b=18}$=0.95 pu).}

\textcolor[rgb]{0,0,0}{According to Fig. \ref{DERPg}, from period 9 to 13, the load increases and the production of \textsc{der} is raised to avoid under-voltages.
Especially, $P_g^{der}$ jumps from 191 to 423 kW to avoid under-voltage issues due to the upstream voltage drop shown in Fig. \ref{TapSt} (a).
From period 13 to 15, $P_g^{der}$ is slightly decreased since the load level slightly decreases. $P_g^{der}$ increases again in period 16, as does the load level.
At period 17, the upstream voltage drops for the second time and the load level widely increases.
It is not now possible to hold the voltages within the limits without tap-changing operations.}
Fig. \ref{TapSt} (b) shows the tap-changing operations for all periods.
Instead of 1 step, the tap position is reduced by 2 steps to avoid the additional cost of another tap-changing operation for higher load levels.
Finally, the framework proposes to increase the tap position to reduce the voltage level after the upstream voltage is restored to 1 pu.
This reduces the operation cost, since the load levels decrease due to load-to-voltage dependance. \textcolor[rgb]{0,0,0}{This OC reduction for the remaining periods of the current day and early hours of the next day is worth spending the tap-changing cost. With a short scheduling window ($T$), the scheduling algorithm has no vision of the upcoming periods and avoids spending this tap-changing cost. This leads to higher operation cost. In case 2, the effects of such a short scheduling window are further analyzed.}

\textbf{In Case 2}, the \textsc{sh} is considered to be 1 hour, signifying no vision of the system future.
\textcolor[rgb]{0,0,0}{The tap position is reduced at period 17 to cope with voltage reduction at upstream system (see Fig. \ref{TapSt}(b)).
If the scheduling algorithm was able to see only one more period ahead, the tap position would be reduced by 2 steps.
Since in the next period, the load level increases and the tap position should be lowered again.}
The \textsc{cb} step is increased one time to avoid under voltage during the heavy-load periods.
The daily \textsc{oc} is also higher in this case (Table \ref{Cases}) since the tap changing operations are not justified
just for a single period while these control actions were able to reduce the cost for $t$=21:24 \textcolor[rgb]{0,0,0}{as explained in case 1}.

\textbf{Case 3} is designed to show the necessity of considering the \textsc{dcc} in the objective function.
The \textsc{dcc} is just excluded from the objective function. For calculating the total cost in Table \ref{Cases}, the \textsc{dcc} is included. As can be seen in Fig. \ref{TapSt} (b), the tap position is freely changed \textcolor[rgb]{0,0,0}{according to the load pattern, upstream system voltage and $\rho_A$.}
The results shows that if the \textsc{dcc} is not taken into account, a lower \textsc{oc} is achieved. However, the higher \textsc{dcc} increases the total cost.

\textcolor[rgb]{0,0,0}{The interactions between the tap-changing operations and the power output of the \textsc{der} in this study are also interesting.
In order to better understand this interconnections, the load levels and $\rho_A$ (presented in Fig. \ref{LoadCost}) should also be taken into account.
In period 1, $\rho_A$ is relatively high.
The scheduling algorithm takes the opportunity to reduce the system cost, by reducing the voltage at substation level.
To avoid under-voltages at bus 18, $P_g^{der}$ is increased. $\rho_A$ is even lower than $\rho_{der}$=60 (\euro/MWh). However, the load reduction due to increasing the tap position is worth spending more money on supplying a part of the system demand by the \textsc{der} in this period.
For period 2, the average $\rho_A$ decreases and supplying a part of the system demand by the \textsc{der} is not justified.
For periods 3-7 the load level is lower and therefore, with a low $P_g^{der}$ the under-voltage at bus 18 is avoided. Increasing $P_g^{der}$ increases the \textsc{oc}, but the load reduction due to the tap increase justifies the solutions shown in Figs. \ref{TapSt} and \ref{DERPg} for these periods.
With higher $\rho_A$ and lower $P_d$ in period 8, the tap position is increased by two steps and $P_g^{der}$ is increased again to avoid under-voltages.}

\textbf{Case 4}: In case 1, relatively high $D_{OLTC}$ and $D_{CB}$ were deliberately selected (considering 3 tap changing operations per day, $D_{OLTC}$=\euro20, is equivalent to maintenance cost of 660 (\euro1000/year) for the \textsc{oltc} transformers). To show that with the higher prices for \textsc{dca}s, the proposed algorithm is still able to reduce the cost, the simulations are repeated with $D_{OLTC}$=\euro40 and $D_{CB}$=\euro20, and the exact same results were found.
\textcolor[rgb]{0,0,0}{The only difference between cases 1 and 4 is on their \textsc{dcc} (Table \ref{Cases}) due to the different discrete control prices.}

\textbf{In Case 5}, the load-to-voltage dependence is neglected.
\textcolor[rgb]{0,0,0}{This means $a'$ is set to 0 in the scheduling algorithm. However, the \textsc{oc} is calculated in Table \ref{Cases} with the real values of the load model parameters.
With $a'$=0, the load levels are assumed to be independent of the voltage levels.
Therefore,} the algorithm increases the voltage levels to reduce the load current and network losses.
The tap position and \textsc{cb} step are set to the lowest and highest values, respectively. Though this leads to the lower \textsc{dcc}, the system \textsc{oc} and total cost are higher according to Table \ref{Cases}.
As shown in Fig. \ref{CostLoss}, even the loss reduction is not realized in reality by this inaccurate load model. 

\subsection{Effects of Inaccurate Upstream System Models}\label{UpSys}
This subsection discusses what happens in case of applying an inaccurate upstream model. One option is to consider the voltage of the primary side as the Thevenin voltage neglecting the Thevenin impedance. This study is referred to as Case 0.
To extract the results for Case 0, at the start of each period, $V_p$ is calculated considering a $Z_{th}$
equal to the one considered for Cases 1-5 and the Thevenin voltage shown in Fig.
\ref{TapSt}(b). This step is designed to find the values of $V_p$. \textcolor[rgb]{0,0,0}{In reality these values} are found through the measurements.
Then the proposed algorithm is applied assuming $Z_{th}$=0 and $V_{th}$=$V_p$.
When the solution is found, the line currents, bus voltages, power losses and system cost are calculated assuming a $Z_{th}$ equal to the one considered for Cases 1-5.
This approach is applied for all 24 hours of the current day with $T$=20 hours.
Table \ref{Cases} shows that the system cost would be about 2\% higher comparing to Case 1.
To elaborate, it should be noted that with $Z_{th}$=0, $V_p$ would remain fixed as the control parameters are changed in downstream network. However, in reality, by applying the changes, the primary current varies, and sometimes the actual value of $V_p$ is lower/higher than the assumed fixed value and therefore, due to the voltage dependent characteristics of the system loads, the demand would be lower/higher. The control plans at different periods are also different from Case 1.
\begin{figure}[t!]
  \centering
\includegraphics[clip = true,width=0.95\columnwidth,trim=1.1cm 0.09cm 1.2cm 0.3cm,]{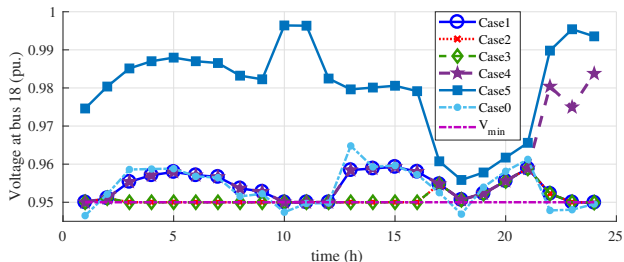}
\vspace{-3.5mm}
  \caption{Voltage level at bus 18 for $\tau=0$ and periods 1-24.}
	\vspace{-1.5mm}
\label{Voltage2}
\vspace{-1.5mm}
\end{figure}

This simplified upstream model also leads to voltage constraints violations. According to Fig. \ref{Voltage1}, bus 18 has the lowest voltage level.
Fig. \ref{Voltage2} shows the 24-hour voltage profile at this bus for all cases in the most probable scenario.
In Cases 1-5, there is no under-voltage issue. The under-voltage issue observed in Case 0 shows that without an accurate upstream model, the solutions can be infeasible. The upstream models should also be kept up-to-date.
\textcolor[rgb]{0,0,0}{Another important observation is that based on Fig. \ref{Voltage2}, }for Case 0 and Cases 1-4, the proposed scheduling framework has tried to keep the voltage level as low as possible to reduce the system demand and system cost. In Case 5, the voltage levels at this bus are higher due to neglecting the voltage dependent nature of the system loads.
\subsection{Effects of Load Characteristics}\label{LoadModelEffect}
\begin{figure}[t!]
  \centering
\includegraphics[clip = true,width=0.95\columnwidth]{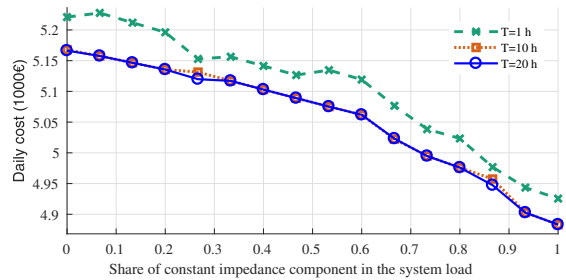}
\vspace{-2mm}
  \caption{System daily cost as a function of constant impedance share.}
\label{Load2Vol}
\vspace{-2mm}
\end{figure}

Fig. \ref{Load2Vol} shows the daily costs, i.e, the sum of system costs \textcolor[rgb]{0,0,0}{for the first hour of all }24 scheduling windows \textcolor[rgb]{0,0,0}{for different values of} the constant impedance share ($a'$) of system load.
$a'=0$ indicates a fully constant power load.
Comparison of these costs for $T$=1, 10, 20 h,
gives an insight into the importance of applying the proposed method based on \textsc{rhc}.
The results of previous sections show that the voltage-dependent nature of loads can be deemed
as an opportunity to reduce the operation cost. With $T$=1, in some cases, the system costs increase as $a'$ increases.
The reason lies in the fact that with $T$=1, the scheduling framework is unable to take the future operation conditions
into account and makes greedy decisions for the early stages which condemn moving towards the optimal operation conditions in the upcoming periods
since the required changes are not justified due to their high cost.

As $a'$ increases, the cost reduction is steeper for some values. \textcolor[rgb]{0,0,0}{These steeper cost reductions happen due to cost-reducing \textsc{dca}s.
To elaborate, for some values of $a'$, cost-reducing \textsc{dca}s might be available, but it might not be possible to apply them due to constraints violations. Increasing the tap position (which reduces the voltage levels) can be considered as one of these \textsc{dca}s. For the higher values of $a'$, after applying these cost-reducing \textsc{dca}s, voltage levels are be lower. These lower voltages reduce the active and reactive power demands due to the voltage-dependent nature of loads. Since $a'$ is relatively high, these demand reductions are considerable. These demand reductions increases the voltage levels so that at a lower demand level, the bus voltages are acceptable according to the standard.}
\subsection{\textcolor[rgb]{0,0,0}{Performance of the Proposed Solution Methodology}}\label{SolTime}
\begin{figure}[!t]
\centering
\subfloat[Solution time for each scheduling window]{\includegraphics[width=0.735\columnwidth,trim=0.3cm 0.05cm 0.7cm 0.2cm,clip=true]{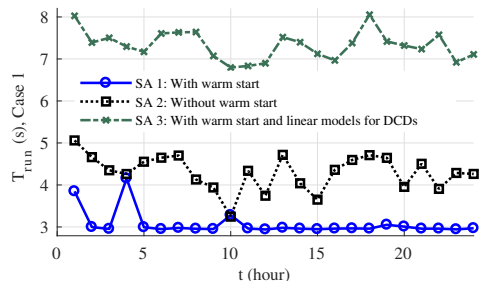}
\label{SolTa}}
\hfill
\subfloat[Average solution time]{\includegraphics[width=0.435\columnwidth,trim=6.65cm 7.9cm 17cm 2.60cm,clip=true]{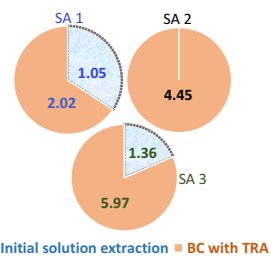}
\label{SolTb}}
\vspace{-1mm}
\caption{Solution times in Case 1 with different solution approaches.}
\vspace{-3mm}
\label{SolT}
\end{figure}

The solution times of each scheduling windows ($T$=20 h) under \textsc{rhc} technique are presented in Fig. \ref{SolTa}
for three different solution approaches (\textsc{sa}s). In SA 1, an initial solution is first found using the method described in \ref{Sol}.
Compared to SA 2, which solves the problem using proposed method without this warm start, the solution time under SA 1 is almost always lower.
By giving a warm start to the proposed solution method,
the average reduction in solution time is more than 30 \%.
Instead of the perturbed models of \textsc{dcd}s in SA 1 and SA 2,
SA 3 models these devices using a set of originally linear equations based on auxiliary binary variables (see subsection \ref{Slow_Control}).
According to Fig. \ref{SolTa}, the solution time is always higher with SA 3,
since the number of integer variables and branching burden are profoundly higher.
Fig. \ref{SolTb} shows the average solution time for all \textsc{sa}s.
\textcolor[rgb]{0,0,0}{Applying the proposed solution methodology (\textsc{sa}1), the solution times are quite acceptable. These study shows the performance of the proposed method and justified the application of the proposed \textsc{bc}-based solution methodology in terms of solution speed.}

\begin{figure}[t!]
  \centering
\includegraphics[clip = true,width=0.95\columnwidth,trim=1.2cm 0.05cm 1.55cm 0.60cm,]{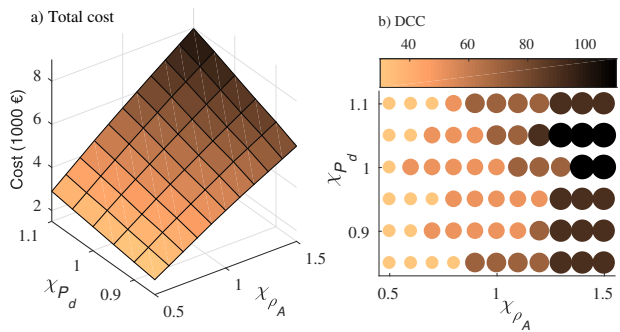}
\vspace{-3mm}
  \caption{\textcolor[rgb]{0,0,0}{a) Total daily cost and b) daily \textsc{dcc}. $\chi_{\rho_A}$: the scaling factor of hourly energy purchase prices. $\chi_{P_d}$: the scaling factor of power demands.}}
\vspace{-2mm}
\label{PLS}
\vspace{-2.5mm}
\end{figure}

\textcolor[rgb]{0,0,0}{Concerning the solution speed, the performance of \textsc{bc} technique has been proven in the literature \cite{Kronqvist2019}. Moreover, the heuristic expediting techniques, such as the one introduced in subsection \ref{Sol}, can be applied to find a tight upper bound on the problem objective function. These approaches are very effective in combination with \textsc{bc} technique, since they widely reduce the branching burden.
This is the distinguished advantage of \textsc{bc} technique in terms of solution speed.}

\textcolor[rgb]{0,0,0}{In terms of solution optimality, \textsc{bc} technique accurately deals with the problem discrete optimization variables.
A \textsc{tra} was adopted in this paper to solve the integer relaxed problems during the branching process.
Byrd-Omojokun technique \cite{BAHRAMI2020} was applied to cope with the inconsistency between the problem constraints and step-size constraints.
It is the ability of \textsc{tra}s to deal with this inconsistency that makes them globally convergent algorithms, compared to the other sequential approaches \cite{BAHRAMI2020}.
The performance of \textsc{bc} technique and the adopted \textsc{tra} were theoretically analyzed in the literature. Here, the overall performance of the proposed methodology is analyzed based on the sensitivity studies conducted in this paper.}

\textcolor[rgb]{0,0,0}{In cases 1-5, the results of applying the proposed scheduling algorithm for the different values of $D_{OLTC}$, $D_{CB}$ and $T$ were presented.
The sensitivity of the results to the change in these input parameters was analyzed.
The rationales behind the variation in the optimal solution and the value of objective function due to the change in these parameters were presented.
For all cases 1-5, the proposed method were applied 24 times to solve the \textsc{dsp} in 24 $T$-hour scheduling windows. Each scheduling window has its own $T$-hour load and average energy purchase price profiles. The variations of the optimal solutions for successive scheduling windows were tracked and it was shown that these variations are rational and were expected.
In subsection \ref{LoadModelEffect}, the sensitivity of the solution to the constant impedance coefficient of the load model was analyzed. The rationales behind the changes in the results were also presented.}

\textcolor[rgb]{0,0,0}{Another sensitivity study is conducted here to analyze the changes of the optimal solution due to the change in the load levels and energy prices. The values $\rho_A$ and load factor for all 24 hours (see Fig. \ref{LoadCost} (a)) are multiplied by scaling factors $\chi_{\rho_A}$ and $\chi_{P_d}$, respectively. $\chi_{\rho_A}$ is changed from 0.5 to 1.5. $\chi_{P_d}$ is changed from 0.85 to 1.1. The proposed algorithm is applied 24 times for each pair of $\chi_{\rho_A}$-$\chi_{P_d}$, with $D_{OLTC}$=\euro20, $D_{CB}$=\euro10 and $T$=20 h. Fig. \ref{PLS} (a) shows as the scaling factors change, the total scheduling cost changes monotonically and smoothly. As $\chi_{\rho_A}$ increases the total cost increases. The slope of such increase is always higher for the higher values of $\chi_{P_d}$. As $\chi_{P_d}$ increases the total cost increases. The slope of such increase is always higher for the higher values of $\chi_{\rho_A}$. This is an indicator of the overall robustness of the algorithm. Fig. \ref{PLS} (b) shows the \textsc{bcc}s in this study. The larger and darker circles indicate higher \textsc{dcd}s.
According to \ref{PLS} (b), as $\chi_{\rho_A}$ increases, the \textsc{dcc} increases or does not change. The reason is that the higher energy prices justify more \textsc{dca}s.
As $\chi_{P_d}$ changes, the \textsc{dcc} does not vary monotonically.
Despite the non-monotonic behavior of \textsc{dcc}, the total cost monotonically changes as discussed earlier. This best indicates the interactions between the \textsc{dca}s, fast control actions and load level and confirms the results of cases 1-5.
The results of cases 1-5, subsection \ref{LoadModelEffect} and this subsection indicate the robustness of the proposed scheduling algorithm against the variations of the problem input parameters.}
\section{Conclusions}\label{conclusion}
The proposed solution methodology is able to solve the stochastic \textsc{dsp} within an acceptable solution time.
The \textsc{rhc} technique is able to effectively co-optimize the operation of discrete and continuous control devices.
\textcolor[rgb]{0,0,0}{With a short scheduling window ($T$), the discrete control actions are not justified unless necessary.
The scheduling algorithm cannot see the effects of \textsc{dca}s in reducing the \textsc{oc} of the upcoming periods.
Therefore, the \textsc{dca}s might not be justified due to their higher cost. This results in the higher total costs.
This means with a short \textsc{sh} it is not possible to exploit the voltage dependence nature of loads to reduce the system cost.}
With a proper tap-changing strategy in a longer scheduling window, lower costs can be achieved.
\textcolor[rgb]{0,0,0}{The continuous control actions widely affect the required discrete control actions, and vice versa. It is imperative to simultaneously schedule these devices to take the interactions between these actions into account. This helps to avoid infeasible or sub-optimal solutions.}
It has also been observed that without updating the loads and upstream system models, the solution will be suboptimal or even infeasible.
The originally linear models (based on auxiliary binary variables) for \textsc{dcd}s profoundly increase the solution time compared to the perturbed models. \textcolor[rgb]{0,0,0}{The reason is the huge branching burden required to handle these auxiliary binary variables.}

\textcolor[rgb]{0,0,0}{To exploits the advantages of both short and long \textsc{sh}s, further studies are required to extend the proposed \textsc{rhc}-based scheduling algorithm.
Similar to the operation of \textsc{dcd}s, the operation of storage systems and most demand response programs cannot be modeled within a short \textsc{sh}.
The inclusion of the operation of storage systems and demand response programs in the proposed scheduling algorithm is suggested for future research activities on this topic.}

\bibliographystyle{IEEEtran}
\bibliography{ref}
\begin{IEEEbiographynophoto}{Alireza Nouri}
(M’17) received the Ph.D. degree in electrical engineering from the Sharif University of Technology, Tehran, Iran, in 2016. He is now a senior power system researcher with the School of Electrical and Electronic Engineering, University College Dublin. His current research has been focused on power systems optimization and control.
\end{IEEEbiographynophoto}

\begin{IEEEbiographynophoto}{Alireza Soroudi}
(M’13–SM’16) received the Ph.D. degree in electrical engineering from the Grenoble-INP, Grenoble, France, in 2012. He is an Assistant Professor at UCD. His research interests include power systems planning and operation, risk, and uncertainty modeling.
\end{IEEEbiographynophoto}

\begin{IEEEbiographynophoto}{Andrew Keane}
(S’04–M’07–SM’14) received the Ph.D. degree in electrical engineering from the University College Dublin (UCD), Dublin, Ireland, in 2007. He is a Professor and Director of the Energy Institute at UCD. His research interests include power systems planning and operation, distributed energy resources, and distribution networks.
\end{IEEEbiographynophoto}

\end{document}